\documentclass[10pt,twocolumn,letterpaper]{article}

\usepackage{cvpr}              % To produce the CAMERA-READY version
\usepackage[accsupp]{axessibility}

\usepackage{graphicx}
\usepackage{amsmath}
\usepackage{amssymb}
\usepackage{booktabs}
\usepackage{multirow}
\usepackage{epstopdf}
\usepackage{epsfig}
\usepackage{caption,subcaption}

\usepackage{amssymb,amsfonts, mathtools}
\usepackage{diagbox} % 
\usepackage{algorithm}
\usepackage{bm}
\usepackage[noend]{algpseudocode}

\usepackage{enumitem}

\usepackage{pifont}

\usepackage{amsthm}

\theoremstyle{definition}

\usepackage[pagebackref,breaklinks,colorlinks]{hyperref}

\usepackage[capitalize]{cleveref}
\crefname{section}{Sec.}{Secs.}
\Crefname{section}{Section}{Sections}
\Crefname{table}{Table}{Tables}
\crefname{table}{Tab.}{Tabs.}

\def\cvprPaperID{10112} % *** Enter the CVPR Paper ID here
\def\confName{CVPR}
\def\confYear{2022}

\begin{document}

%%%%%%%%% TITLE - PLEASE UPDATE
% \title{Layer-wised Model Aggregation for Personalized Federated Learning}

\title{Layer-wised Model Aggregation for Personalized Federated Learning}

\author{%
  Xiaosong~Ma$^{\dagger,1}$,
  Jie~Zhang$^{\dagger,1}$,
  Song~Guo$^{\ast,1,2}$,
%   \thanks{Corresponding author}~~,
%   \textsuperscript{1,\thanks{Corresponding author}}~, 
  and~Wenchao~Xu$^{1}$\\
%   \textsuperscript{1}
  \textsuperscript{1}Department of Computing, The Hong Kong Polytechnic University\\
  \textsuperscript{2}The Hong Kong Polytechnic University Shenzhen Research Institute\\
  \texttt{jieaa.zhang@connect.polyu.hk}, \texttt{maxiaosong16@gmail.com},  \\
  \texttt{\{song.guo,wenchao.xu\}@polyu.edu.hk}
}
% \author{First Author\\
% Institution1\\
% Institution1 address\\
% {\tt\small firstauthor@i1.org}
% % For a paper whose authors are all at the same institution,
% % omit the following lines up until the closing ``}''.
% % Additional authors and addresses can be added with ``\and'',
% % just like the second author.
% % To save space, use either the email address or home page, not both
% \and
% Second Author\\
% Institution2\\
% First line of institution2 address\\
% {\tt\small secondauthor@i2.org}
% }
\maketitle

\def\thefootnote{$\dagger$}\footnotetext{Equal contribution}
\def\thefootnote{$\ast$}\footnotetext{Corresponding author}

\pagestyle{empty}  % no page number for the second and the later pages
\thispagestyle{empty} % no page number for the first page

%%%%%%%%% ABSTRACT
% 1. pFL
% 2. Challenges
% 3. Methodology
% 4. Evaluation
% 
\begin{abstract}
\noindent
Personalized Federated Learning (pFL) not only can capture the common priors from broad range of distributed data, but also support customized models for heterogeneous clients. Researches over the past few years have applied the weighted aggregation manner to produce personalized models, where the weights are determined by calibrating the distance of the entire model parameters or loss values, and have yet to consider the layer-level impacts to the aggregation process, leading to lagged model convergence and inadequate personalization over non-IID datasets. In this paper, we propose a novel pFL training framework dubbed Layer-wised Personalized Federated learning (pFedLA) that can discern the importance of each layer from different clients, and thus is able to optimize the personalized model aggregation for clients with heterogeneous data. Specifically, we employ a dedicated hypernetwork per client on the server side, which is trained to identify the mutual contribution factors at layer granularity. Meanwhile, a parameterized mechanism is introduced to update the layer-wised aggregation weights to progressively exploit the inter-user similarity and realize accurate model personalization. %
% brand-new parameters sharing mechanism among different clients that can effectively realize the model personalization.
% In addition, to further reduce the communication overhead for pFL clients at scale, we also investigated   
% Furthermore, to improve the communication efficiency for distributed clients at scale, we provide an improved version of pFedLA to aggregate only the top $K$ layers with highest weights. 
Extensive experiments are conducted over different models and learning tasks, and we show that the proposed methods achieve significantly higher performance than state-of-the-art pFL methods.
\end{abstract} 

\section{Introduction}
\label{sec:intro}

% \begin{figure}[t]
% \centering
%     \includegraphics[width=0.4\textwidth]{ pics/layer-wised vs. model-wised.pdf}
%     \caption{Layer-wised vs. Model-wised}
%     \label{fig:1}
% \end{figure}

\begin{figure}[t]
\centering
\subcaptionbox{\label{1}}{\includegraphics[width = 0.455\linewidth]{ 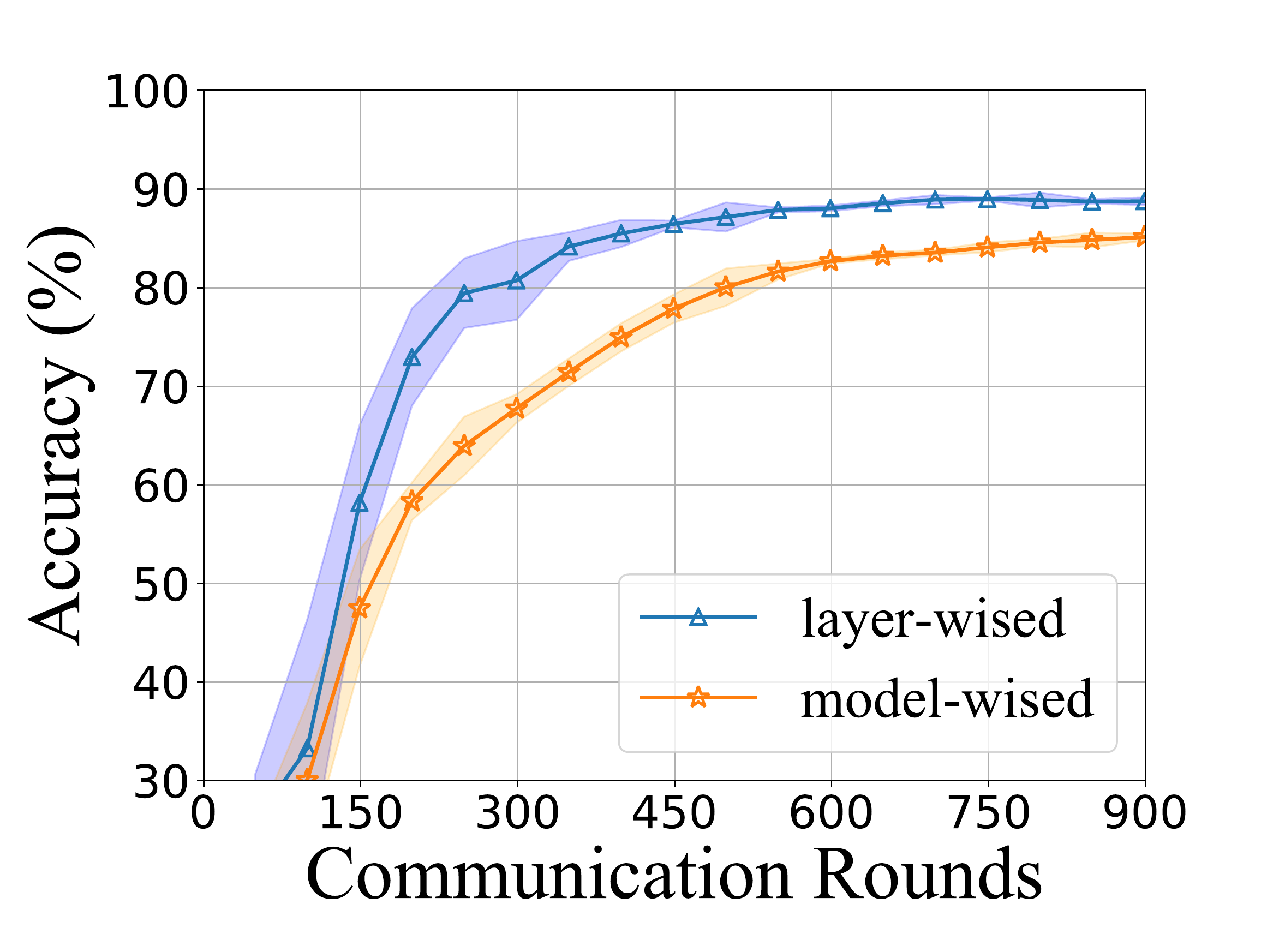}}
\subcaptionbox{\label{2}}{\includegraphics[width = 0.535\linewidth]{ 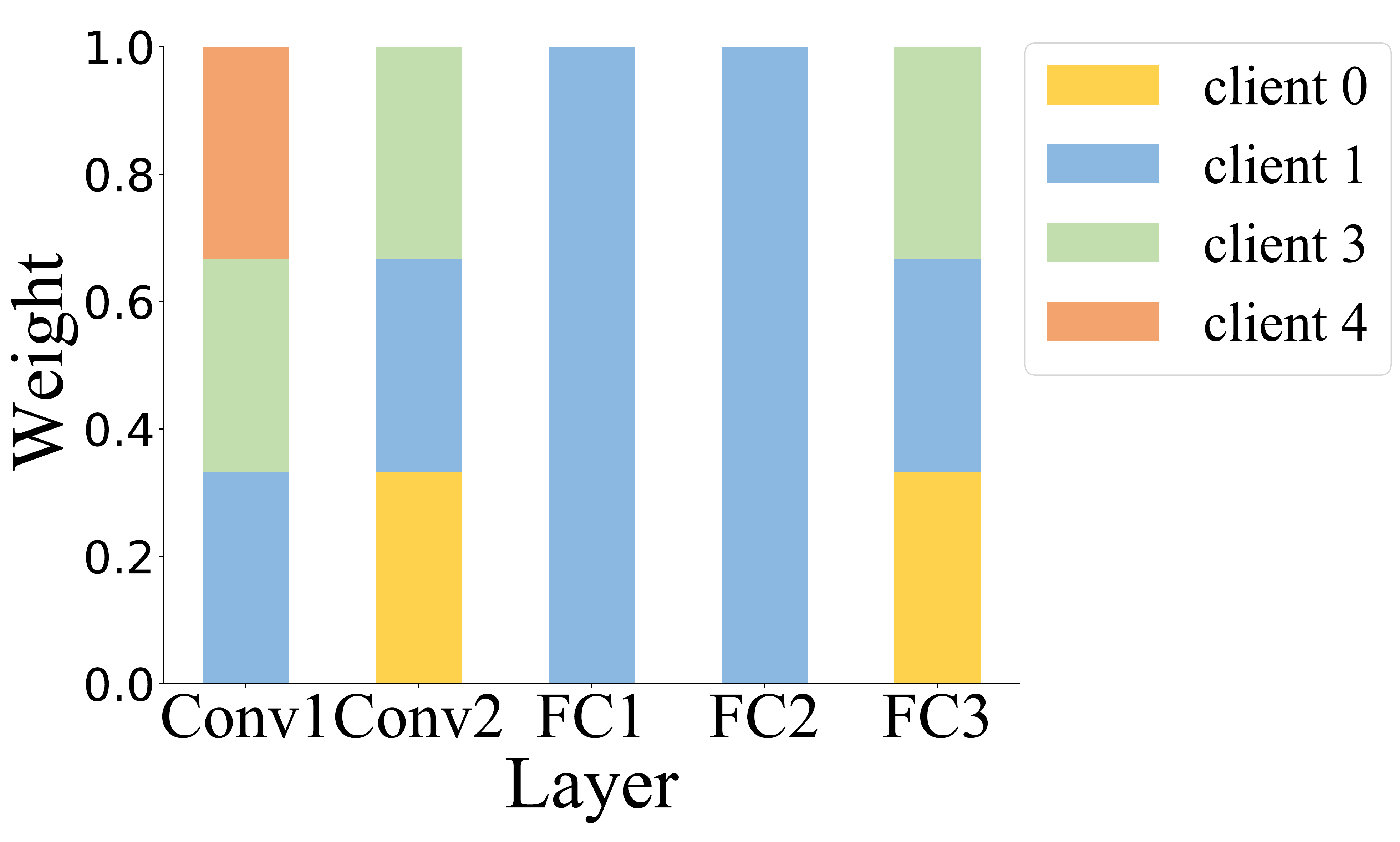}}
\caption{A toy example: Layer-wised vs. Model-wised aggregation method. (a) Model performance of client 1. Both of two methods perform similarity-based personalized aggregation. i.e., layer-wised: perform personalized aggregation by calculating the similarity between layers; model-wised: perform personalized aggregation by calculating the similarity between models. (b) The weight of each layer for client 1 in the last communication round.}
\label{fig:1}
\vspace{-0.3cm}
\end{figure}

Federated learning (FL) has emerged as a prominent collaborative machine learning framework to exploit inter-user similarities without sharing the private data \cite{DBLP:conf/aistats/McMahanMRHA17,tran2019federated,zhang2021adaptive}. 
% Federated Learning (FL) \cite{DBLP:conf/aistats/McMahanMRHA17} has emerged as an efficient paradigm that collaboratively trains a shared machine learning model among multiple clients without exposing their local raw data.  （wenchao：这段话好像跟NIPS开头重复了）
%
%For large inter-user distances, i.e., the datasets are non-IID (independent and identically distributed) among clients, sharing a global model for all of them may lead to   
When users' datasets are non-IID (independent and identically distributed), i.e., the inter-user distances are large \cite{li2018federated,zhang2021edge}, sharing a global model for all clients may lead to slow convergence or poor inference performance as the model may significantly deviate from their local data \cite{hsieh2020non,zhao2018federated}.   
% One of the biggest challenges for FL is the issue of data heterogeneity, that is, the data is non-IID across different clients \cite{li2018federated,mcmahan14advances}. In such settings, learning a single global model for all clients (e.g., by minimizing average loss) can result in significant bias and poor performance \cite{hsieh2020non,zhao2018federated}. 
%

To deal with such statistical diversity, personalized federated learning (pFL) mechanisms are proposed to allow each client to train a customized model to adapt to their own data distribution \cite{hanzely2020lower,DBLP:conf/nips/0001MO20,li2021ditto,huang2021personalized}. Literature status quo to achieve pFL include the data-based approaches, i.e., smoothing the statistical heterogeneity among clients’ datasets \cite{jeong2018communication,duan2020self}, the single-model approaches, e.g., regularization \cite{li2021ditto,t2020personalized}, meta-learning \cite{DBLP:conf/nips/0001MO20}, parameter decoupling \cite{DBLP:conf/icml/CollinsHMS21,liang2020think,DBLP:conf/iclr/LiJZKD21}, and the multiple-model ways, i.e., train personalized models for each client \cite{zhang2020personalized,huang2021personalized}, which can produce personalized models for each client via weighted combinations of clients' models.         
% Apart from the data-based pFL approaches that focus on smoothing the statistical heterogeneity among clients’ datasets \cite{DBLP:journals/corr/abs-1811-11479,duan2020self}, and the single model-based pFL methods (i.e., regularization \cite{li2021ditto,t2020personalized}, meta-learning \cite{DBLP:conf/nips/0001MO20}, parameter decoupling \cite{DBLP:conf/icml/CollinsHMS21,liang2020think}), 
% another promising approach is to train multiple unique personalized models for clients in the federation \cite{zhang2020personalized,huang2021personalized}. In this method, the key point is to compute a weighted combination of available models to best align with each client's target.
% These approaches are well-suited for applications with diverse data distributions, or when the heterogeneity of the underlying data distributions is unclear. 
Existing pFL methods apply a distance metric among the whole model parameters or loss values of different clients, which is insufficient to exploit their heterogeneity since the overall distance metric cannot always reflect the importance of each local model and can lead to inaccurate combining weights or unbalance contribution from non-IID distributed datasets, and thus prevent further personalization for clients at scale. The main reason is that different layers of a neural network can have different utilities, e.g., the shallow layers focus more on local feature extraction, while the deeper layers are for extracting global features \cite{xu2018attention,yu2018deep,li2018improved,cramer2020chirping,lee2021layer}. Measuring the model distances would ignore such layer-level differences, and cause inaccurate personalization that hinders the pFL training efficiency.  

% rely on a distance metric (i.e., model parameter, loss) between different clients, which may not accurately reflect the importance of each local model. 
%
% Besides, as different layers of neural network can have a different purpose, e.g., the shallow layer focuses more on local feature extraction, while the deeper layer extracts more global features \cite{xu2018attention,yu2018deep,li2018improved,cramer2020chirping}, considering a equivalent contribution for each layer in personalized aggregation phase hinders the further improvement of training performance.
%
% To solve above challenges, 
In this paper, we propose a band-new pFL framework that can realize the layer-level aggregation for FL personalization, which can accurately recognize the utility of each layer from clients' model for adequate personalization, and thus can improve the training performance over non-IID datasets. A toy example is presented to illustrate that traditional model-level aggregation based pFL method fails in reflecting the inner relationship among all local models, which motivates us to exploit an effective way to discern the layer-level impacts during the pFL training procedure.         

% In this paper, we aim to develop a new federated training framework, which allows for layer-wised personalized model aggregation. ppp
% To illustrate the idea, we first present a toy example that illustrates how traditional model-wised aggregation may not be an optimal choice to measure the inner relationship between the local models in pFL.
% In this paper, we aim to develop a new approach to effectively measure the importance of the local model parameters.
% To illustrate the idea, we first present a toy example that illustrates how traditional distance-based metric may not be an optimal choice to measure the inner relationship between the local models in pFL.

%-------------------------------------------------------------------------

% Toy example 需要重新设计：
% 1. 体现layer-wise vs. model-wise 的好处
% 2. 体现similarity-based measurement 的不足之处，i.e., not optimal
\noindent
% \textbf{Observation of Layer-wised Personalized Aggregation in a FL Toy Example.}
\textbf{Observation of Layer-wised Personalized Aggregation.}
In the toy example, we consider six clients to collaboratively learn their personalized models for a nine-class classification task. The average model accuracy is obtained via both the layer-wised and model-wised aggregation approaches, which utilize the inter-layer and inter-model similarities respectively. Figure~\ref{fig:1} shows that higher model accuracy can be achieved by the layer-wised approach comparing with the model-wised one for a certain client. The weights of layers for this client after the last communication round are also plotted, and we show that applying different weights for different layers, e.g., the first and second fully-connected layer (i.e., FC1, FC2) on client 1 have larger weights, while the second convolution layer, i.e., Conv1 layer has smaller weights, can produce significant performance gain for the personalized model accuracy.   

The toy example demonstrates the potential of the layer-wised aggregation to achieve higher performance than traditional model based pFL methods, since the layer-level similarities can reflect more accurate correlation among clients. By exploiting such layer-wised similarity and identifying the layer-level inter-user contribution, it is promising to produce efficient and effective personalized models for all clients. Motivated by such observation, we propose a novel federated training framework, namely, pFedLA, which adaptively facilitates the underlying collaboration between clients in a layer-wised manner. 
%% JIe
Specifically, at the server side, we introduce a dedicated hypernetwork for each client to learn the weights of cross-clients' layers during the pFL training procedure, which is shown to effectively boost the personalization over non-IID datasets. 
Extensive experiments are conducted, and we demonstrate that the proposed pFedLA can achieve higher performance than the state-of-the-art baselines over widely used models and datasets, i.e., EMNIST, FashionMNIST, CIFAR10 and CIFAR100.
% From the motivation example, we observe that traditional model-based pFL methods cannot accurately reflect the correlation between different clients, which may lead to deviations from the best optimization direction of the objective function. 
% Moreover, different layers of neural network can have different contribution to the personalized model aggregation phase.
% In contact, 
%
% Motivated by the above insight, this paper proposes a novel federated training framework, called pFedLA, which adaptively facilitates the underlying collaboration between clients in a layer-wised manner. 
%
% uses hypernetworks to automatically generate and allows for layer-wised personalized model aggregation. 
% More specifically, by training multiple hypernetworks in the server,  the aggregation weights of each layer from different clients can be automatically generated, which allows for a accurate evaluation on a personalized target objective.
% %
% The proposed architecture achieves state-of-the-art performance on the most widely used benchmarks for
% various image classfication tasks: EMNIST, FashionMNIST, CIFAR10 and CIFAR100. 
%
The contributions of the paper are summarized as follows:
\begin{itemize}
    \item To the best of our knowledge, this paper is the first to explicitly reveal the benefits of layer-wised aggregation comparing with model-wised approaches in pFL among heterogeneous FL clients;
    
    % this is the first work to explicitly reveal that distance-based metrics cannot well-reflect the contributions of local models to the personalized aggregation phase. 
    % In addition, we 
    % We propose a  
    \item We propose a layer-wised personalized federated learning (pFedLA) training framework that can effectively exploit the inter-user similarities among clients with non-IID data and produce accurate personalized models;
    
    % A method called layer-wised personalized fedearted learning (pFedLA) is proposed which enables effective collaborations for diverse clients in a privacy-preserving manner.
    % \item To achieve  
    \item We conduct extensive experiments on four typical image classification tasks, which demonstrated the superior performance of pFedLA over the state-of-the-art approaches.
    
    % Our achieved superior performance over state-of-the-arts and in-depth analytical experiments demonstrate the efficacy of our approach.
\end{itemize}

\begin{figure*}[t]
\centering
    \includegraphics[width=0.85\textwidth]{ 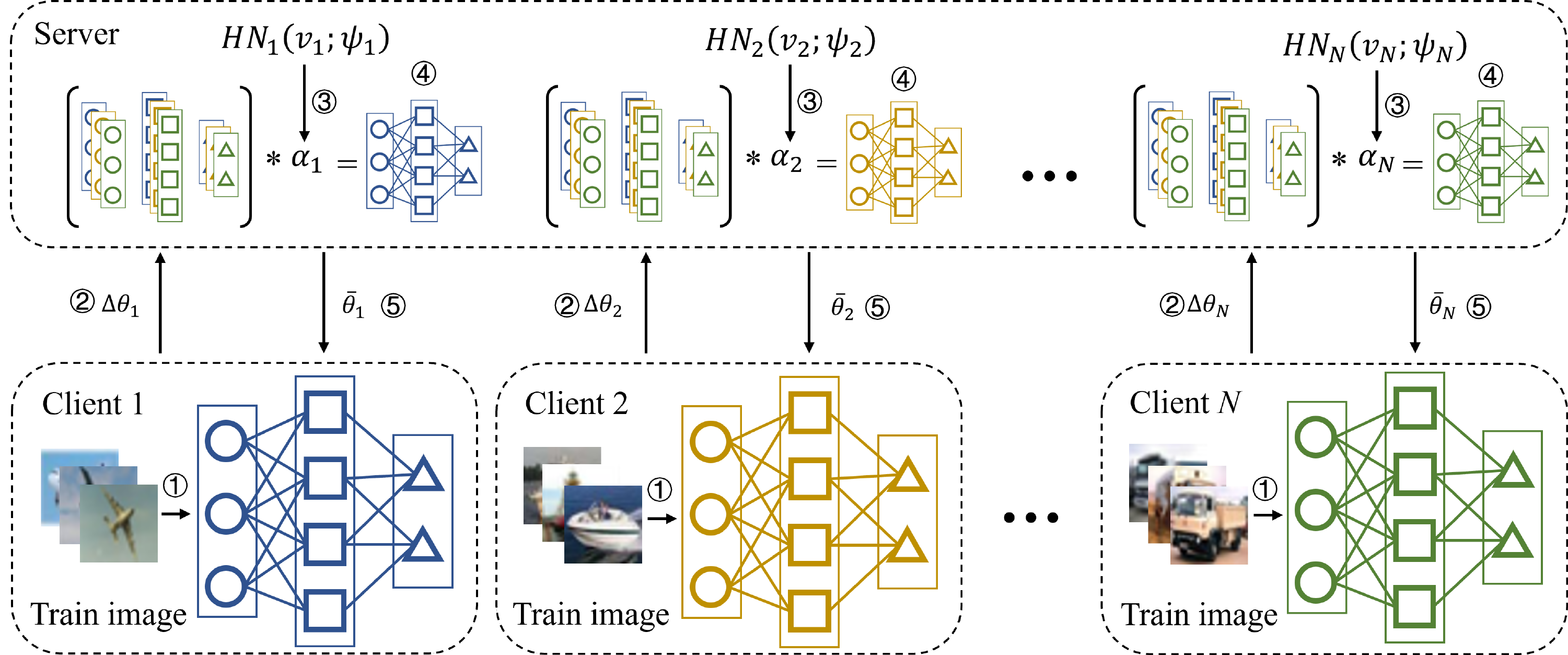}
    \caption{Framework of pFedLA. The workflow contains 5 steps: \ding{172} local training on private data; \ding{173} each client sends the update of parameters $\Delta\theta_i$ to the server; \ding{174} the server updates the aggregation weight matrix $\alpha_i$ by hypernetworks $HN_i(v_i;\psi_i)$ according to $\Delta\theta_i$; \ding{175} the server performs weighted aggregation and outputs personalized model $\bar{\theta}_i$ for the corresponding client; \ding{176} each client downloads the personalized model $\bar{\theta}_i$.}
    \label{fig:framework}
    \vspace{-0.3cm}
\end{figure*}

\section{Related Work}\label{sec:rel}
%------------------------------------------------------------------------

\subsection{Personalized Federated Learning}
Recently, various approaches have been proposed to realize pFL, which can be classified into the data-based and the model-based categories. 
Data-based approaches focus on reducing the statistical heterogeneity among clients’ datasets to boost the model convergence, while model-based approaches emphasize on producing customized model structures or parameters for different clients. 

% enhance the performance of FL models under different levels of personalization.
% two types according to the number of global models applied in the server, i.e., single global model, multiple global models, no global model.

% Data-based pFL methods aim to smooth the divergence of data heterogeneity among different clients. 
The typical way of data-based pFL is to share a small amount
of global data to each client \cite{zhao2018federated}. Jeong et al. \cite{jeong2018communication,duan2020self} focus on data augmentation methods by generating additional data to augment its local data towards yielding an IID dataset. However, these methods usually require the FL server to know the statistical information about clients’ local data distributions (e.g., class sizes, mean and standard deviation), which may potentially violate privacy policy \cite{tan2021towards}. Another line of work considers to design client selection
mechanisms to approach homogeneous data distribution \cite{wang2020optimizing,yang2020federated,lyu2020towards}.

Model-based pFL methods can also be divided into two types: single-model, multiple-model approaches.
Single-model based methods extended from the conventional FL algorithms like FedAvg \cite{DBLP:conf/aistats/McMahanMRHA17} combine the optimization of the local models and global model, which consist of five different kinds of approaches: local fine-tuning \cite{wang2019federated,schneider2019personalization,arivazhagan2019federated}, regularization \cite{t2020personalized,hanzely2020federated,hanzely2020lower}, model mixture
% mixture of global model and local models
\cite{mansour2020three,deng2020adaptive},  meta learning \cite{DBLP:conf/nips/0001MO20,jiang2019improving} and parameter decomposition \cite{bui2019federated,DBLP:conf/icml/CollinsHMS21,arivazhagan2019federated}. %
Considering the diversity and inherent relationship of local data, a multi-model-based approach where multiple global models are trained for heterogeneous clients is more suitable. 
Some researchers \cite{huang2021personalized,DBLP:conf/nips/GhoshCYR20,mansour2020three} propose to train multiple global models at the server, where similar clients are clustered into several groups and different models are trained for each group. 
Another strategy is to collaboratively train a personalized model for each individual client, e.g., FedAMP \cite{huang2021personalized}, FedFomo \cite{zhang2020personalized}, MOCHA \cite{DBLP:conf/nips/SmithCST17}, KT-pFL \cite{zhang2021parameterized} etc.

These literatures treat each client's model as a whole entity, and has yet to consider the layer-wised utility for personalized aggregation. The distance metric for describing the similarity among models is inaccurate and can lead to sub-optimal performance, which motivates us to explore a fine-grained aggregation strategy to adapt to broad range of non-IID clients.

% In all of these methods, each client's model is considered as either a whole entity or a combination of head and network body --- there is no notion of learning a layer-wised personalized aggregation mechanism. Besides, obtaining a personalized global model usually requires to choose a metric (i.e., parameter, loss) to estimate the similarity across different models, which may be biased and sub-optimal. We focus on design a more practical training and aggregation strategy to achieve accurate measurement.
% the different layers of neural network for each client's model is considered as 
% multi-task learning based methods need to solve the data-local quadratic subproblem for every client, which is computationally intensive. It requires all clients to participate in every training round, which is inapplicable for large-scale clients.  
%Besides, it requires all clients participant in training every round since it produces one model per task. 

\subsection{Hypernetworks}
Hypernetworks \cite{ha2016hypernetworks} are used to generate parameters of other neural networks, e.g., a target network, by mapping the embeddings of the target tasks to corresponding model parameters.   
% are networks (also called supernetworks) that generate weights of another network (often referred to as target networks). The main concept is to conduct the relation map between the embedding of the task and the corresponding model parameters. 
% Different from the traditional approaches that use the training set to directly train the model,  hyperNetwork directly output parameters, no need to train, abandon back propagation and gradient descent, and, which is equivalent to HyperNetwork learning how to learn image recognition.
% To put it simply, HyperNetwork is actually a meta network.
% are useful as a modeling tool, and
Hypernetworks have been widely used in various machine learning applications, such as language modeling \cite{suarez2017language,nirkin2021hyperseg}, computer vision \cite{jia2016dynamic,klocek2019hypernetwork,littwin2020infinite}, 3D scene representation \cite{littwin2019deep,sitzmann2020implicit}, hyperparameter optimization \cite{lorraine2018stochastic,li2020dhp,mackay2019self,bae2020delta}, neural architecture search (NAS) \cite{brock2017smash,zhang2018graph}, continual learning \cite{von2019continual} and meta-learning \cite{zhao2020meta}. Shamsian et al. \cite{DBLP:conf/icml/ShamsianNFC21} is the first to apply hypernetworks in FL, which can generate effective personalized model parameters for each client. We show that hypernetworks are capable to evaluate the importance of each model layer, and can boost the personalized aggregation in non-IID scenarios.    
%
% To our knowledge, however, hypernetworks were never proposed for evaluating the importance of model parameters. Therefore, we propose to estimate each layer's contribution to the personalized aggregation phase.

% \section{Problem Formulation and Preliminaries}\label{sec:pre}
%------------------------------------------------------------------------

\section{Method}\label{section:method}
% \section{Analysis}\label{sec:analysis}
In this section, we present the design of the pFedLA framework that applies the hypernetworks to conduct layer-wised personalized aggregation, which is shown in Figure~\ref{fig:framework}.

% start with the formulation for our proposed hypernetworks-based pFL problem.
% We then describe the proposed
% layer-wised personalized federated learning (pFedLA) algorithm to explicitly address the above problem. An overview of the method is shown in Fig.~\ref{fig:framework}.

\subsection{Problem Formulation} 
In pFL, the goal is to collaboratively train personalized models among multiple clients while keeping their local data private. 
% The algorithm we propose, pFedLA, is used to solve the pFL problem. We first formalize a pFL problem: 
Considering $N$ clients with non-IID datasets, let $\mathcal{D}_i=\{(x_{j}^{(i)},y_{j}^{(i)})\}_{i=1}^{m_i}$ ($1 \leq i \leq N$) be the dataset on the $i$-th client, where $x_{j}$ is the $j$-th input data sample, $y_{j}$ is the corresponding label. The size of the datasets on the $i$-th client is denoted by $m_i$. The size of all clients' datasets is $M=\sum_{i=1}^{N}m_i$. Let $\theta_i$ represent the model parameters of client $i$, 
the objective of pFL can be formulated as 
% is obtaining the sequence of parameters:
\begin{equation}
% \begin{aligned}
    % \{\theta_i^*,\dots,\theta_N^*\}
    \Theta^* =\arg \min_{\Theta} \sum_{i=1}^N \frac{m_i}{M} \mathcal{L}_i({\theta_i}),
    \label{eq:conventional_loss}
% \end{aligned}
\end{equation}
where
\begin{equation}
    \mathcal{L}_i({\theta_i}) = \frac{1}{m_i} \sum_{j=1}^{m_i} \mathcal{L}_{CE}({\theta_i}; x_j^{(i)}, y_j^{(i)})
    \label{eq:conventional_local_loss}
\end{equation}
where $\Theta = \{\theta_i,\dots,\theta_N\}$ is the set of personalized parameters for all clients. $\mathcal{L}_i$ is loss function of $i$-th client associated with dataset $\mathcal{D}_i$. The difference between the predicted value and the true label of data samples is measured by $\mathcal{L}_{CE}$, which is the cross-entropy loss. %wenchao: ith associated what?

\subsection{pFedLA Algorithm}\label{subsec:pfedla}
In this section, we present our proposed pFL algorithm pFedLA, which evaluates the importance of each layer from different clients to achieve layer-wised personalized model aggregation. We apply a dedicated hypernetwork for each client on the server and train them to generate aggregation weights for each model layer of different clients. It can be seen from Figure~\ref{fig:framework} that, unlike the general FL framework that generates only one global model, pFedLA maintains a personalized model for each client at the server.
% to better adapt to the data non-IID scenario in FL. 
%
Clients with similar data distribution should have high aggregation weights to reinforce the mutual contribution from each other. Our pFedLA applies a set of aggregation weight matrix $\alpha_i$ at the server side to progressively exploit the inter-user similarities at layer level, which is defined as   

% For each single client, intuitively, clients with similar data distribution have higher value to it. 

% Since the data distribution is different, each other client has a different value. Therefore, a pFL algorithm needs to be able to correctly measure the value of other clients to the target client. 
% pFedLA uses the aggregation weight matrix $\alpha$ to measure the value of other clients to the target client when aggregating the parameters of the personalized model on the server side. The aggregation weight matrix of client $i$, $\alpha_i$, is defined as follows
\begin{equation}
{\alpha_i} = \left[ {\alpha_i^{l1}},{\alpha_i^{l2}},\dots,{\alpha_i^{ln}}\right] =
\left[
  \begin{matrix}
   \alpha_i^{l1,1} & \alpha_i^{l2,1} & \cdots & \alpha_i^{ln,1} \\
   \alpha_i^{l1,2} & \alpha_i^{l2,2} & \cdots & \alpha_i^{ln,2} \\
   \vdots & \vdots & \ddots & \vdots \\
   \alpha_i^{l1,N} & \alpha_i^{l2,N} & \cdots & \alpha_i^{ln,N}
  \end{matrix}
\right]
% \tag{2}
    \label{eq:weight_matrix}
\end{equation}
where $\alpha_i^{ln}$ represents the aggregation weight vector of $n$-th layer in client $i$, while $\alpha_i^{ln,N}$ represents the aggregation weight for client $N$ in $n$-th layer. For all $n$ layers, $\sum_{j=1}^{N} \alpha_i^{ln,j} = 1$.

\begin{figure}[t]
\centering
    \includegraphics[width=0.5\textwidth]{ 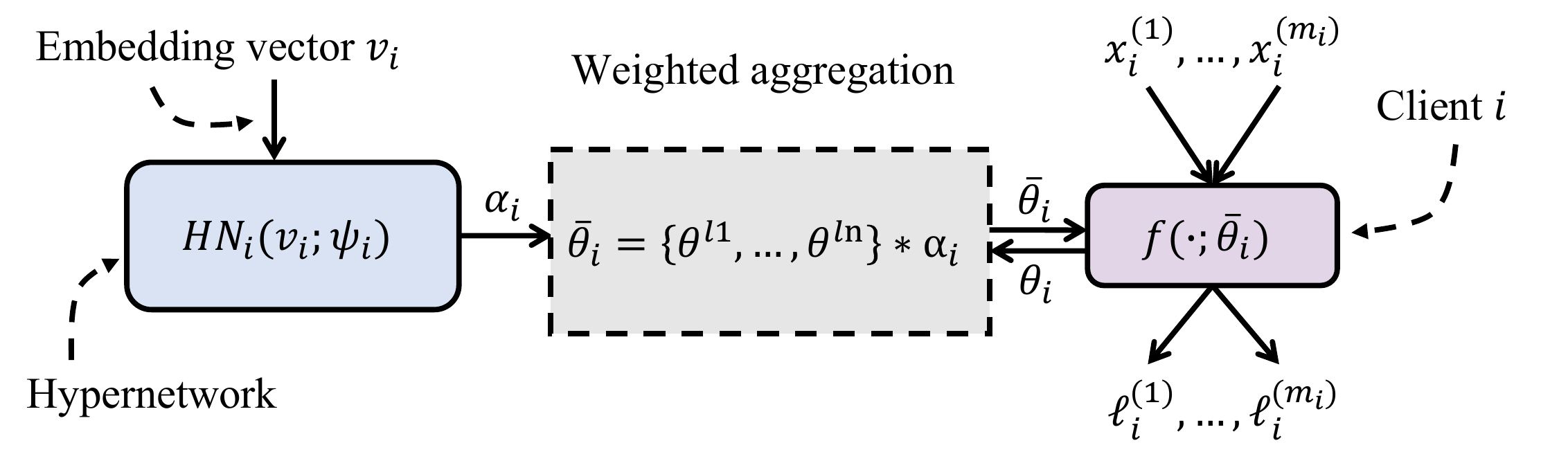}
    \caption{Illustration of one hypernetwork framework used in pFedLA. The hypernetwork $HN_i$ takes the embedding vector $v_i$ as input, and outputs the aggregation weight matrix $\alpha_i$. After the weighted combination with intermediate parameters $\{\theta^{l1}, \dots, \theta^{ln}\}$ and aggregation weight matrix $\alpha_i$, client $i$ can make local training on private data. Note that both $v_i$ and $\psi_i$ are updated during training.}
    \label{fig:hn}
    \vspace{-0.4cm}
\end{figure}

Different with previous pFL algorithms, instead of applying identical weight values for all layers of a client model, pFedLA considers the different utilities of neural layers, and assign a unique weight to each of them to achieve fine-grained personalized aggregation. In addition, unlike traditional methods that mathematically calculate the weights using a distance metric among the entire model parameters \cite{huang2021personalized,zhang2020personalized}, pFedLA parameterized the weights during the training phase via a set of dedicated hypernetworks. The layer-wised weights are determined by the hypernetworks, which are alternatively updated with the personalized model. Such way we can obtain effective weights as their update direction is in line with the optimization direction of the objective function. In the following, we will elaborate the updating process of the aggregation weight matrix $\alpha$ of pFedLA.

\begin{algorithm}[t]
% \small
\caption{pFedLA Algorithm}
\label{alg:1}
\begin{algorithmic}[1]
    \Require dataset $\{\mathcal{D}_1, \mathcal{D}_2,\dots,\mathcal{D}_N\}$, learning rate $\eta$. Total communication rounds $T$.
	\Ensure Trained personalized models $\{\bar{\theta}_1,\bar{\theta}_2,\dots,\bar{\theta}_N\}$.
	\State Initialize the clients' model parameters, hypernetworks parameters and embedding vectors.
% \small
\Procedure{Server executes}  {}
    \For {each communication round $t \in \{1,\dots,T\}$}
    	\For {each client $i$ \textbf{in parallel}}
    		\State $\bar{\theta}_i^{(t+1)}=\{\theta^{l1},\dots,\theta^{ln}\}*{HN_i}(v_i^{(t)};\psi_i^{(t)})$    	    
    		\State $\Delta\theta_i \leftarrow ClientUpdate(\bar{\theta}_i^{(t+1)})$
    	\State Update $\{\theta^{l1},\theta^{l2},\dots,\theta^{ln}\}$ according to $\Delta\theta_i$
    	%\State $v_i^{(t+1)} = v_i^{(t)} - \eta_2[\{\theta^{l1},\theta^{l2},\dots,\theta^{ln}\} * \nabla_{v_i}{HN_i}(v_i^{(t)};\psi_i^{(t)})]^{T}\Delta\theta_i$
    	\State Update $v_i^{(t+1)}$ and $\psi_i^{(t+1)}$ via Eq.~\ref{eq:real_update_of_v}, \ref{eq:real_update_of_psi} 	
    	%\State $\psi_i^{(t+1)} = \psi_i^{(t)} - \eta_2[\{\theta^{l1},\theta^{l2},\dots,\theta^{ln}\} * \nabla_{\psi_i}{HN_i}(v_i^{(t)};\psi_i^{(t)})]^{T}\Delta\theta_i$
    	\EndFor
    \EndFor
\EndProcedure

\Procedure{ClientUpdate} {$\bar{\theta}_i^{(t+1)}$}
    \State Client $i$ receives $\bar{\theta}_i^{(t+1)}$ from the server.
    \State Set $\theta_i = \bar{\theta}_i^{(t+1)}$.
		\For {each local epoch }
    		\For {mini-batch $\xi_t \subseteq \mathcal{D}_i$} 
    		  \State \textbf{Local Training:} $\theta_i = \theta_i - {\eta} \nabla_{\theta_i}\mathcal{L}_{i}(\theta_i;\xi_t)$
    		\EndFor
    	\EndFor
    \Return $\Delta\theta_i = \theta_i - \bar{\theta}_i^{(t+1)}$
\EndProcedure
\end{algorithmic}
\end{algorithm}
% In other words, the update direction of the aggregate weight is the optimization direction of the loss function, which is more direct and effective than using the distance. In the following, we will elaborate the updating process of the aggregation weight matrix $\alpha$ of pFedLA.
% loss function on each client, and are alternatively      

% pFedLA differs from previous pFL algorithm in two points: First, pFedLA optimizes the aggregation weights from the layer-wised level, because the role of each layer in a deep learning model is different, giving an identical weight to all layers on a client is too coarse. The layer-wised aggregation weight can achieve more fine-grained personalized aggregation; 
%
% Second, the aggregation weight in the pFedLA framework, i.e., $\alpha$, is not calculated mathematically by the distance between different clients’ parameters \cite{huang2021personalized,zhang2020personalized},  % (cite articles using distance)
% but is optimized by hypernetworks during the training process. 

% The advantage of using hypernetworks is that the update of the aggregation weights is determined by the update of the loss function on the client, and the training of the local model and the training of the hypernetworks are performed alternately. 

%%
Each hypernetwork consists of several fully connected layers, whose input is an embedding vector that is automatically updated with the model parameters, and the output is the weight matrix $\alpha$. Define the hypernetwork on client $i$ as
\begin{equation}
    {\alpha_i}={HN_i}(v_i;\psi_i),
    \label{eq:HN_i}
\end{equation}
where $v_i$ is the embedding vector and $\psi_i$ is the parameter of client $i$'s hypernetwork (i.e., Figure~\ref{fig:hn}).
Let $\{\theta^{l1}, \theta^{l2}, \dots, \theta^{ln}\}$ be the intermediate parameters of all clients after local training, $\theta^{ln}=\{\theta_1^{ln}, \theta_2^{ln}, \dots, \theta_N^{ln}\}$ is the set of $n$-th layer of all clients, where $\theta_N^{ln}$ are the parameters of $n$-th layer in client $N$. In pFedLA, the model parameters of client $i$ is obtained by weighted aggregation according to $\alpha_i$:
\begin{equation}
% \begin{aligned}
    \bar{\theta}_i=\{\bar{\theta}_i^{l1}, \bar{\theta}_i^{l2}, \dots, \bar{\theta}_i^{ln}\} = \{\theta^{l1}, \theta^{l2}, \dots, \theta^{ln}\} * {\alpha_i},
    \label{eq:theta_i}
% \end{aligned}
\end{equation}
where $\bar{\theta}_i^{ln}$ can also be expressed as: 
\begin{equation}
    \bar{\theta}_i^{ln} = \sum_{j=1}^{N} \theta_j^{ln} \alpha_i^{ln,j}.
    \label{eq:theta_i_n}
\end{equation}
% Insert Eq.3 and Eq.4 into Eq.1, 
Thus the objective function of pFedLA can be derived from Eq.~\ref{eq:conventional_loss} to   
% We adjust the pFL objective (Eq.~\ref{eq:conventional_loss}) according to the above setup to obtain 
% the objective of pFedLA can be obtain:
\begin{equation}
\begin{aligned}
    \arg \min_{V,\Psi}\ \sum_{i=1}^N \frac{m_i}{M} \mathcal{L}_i({\{\theta^{l1}, \theta^{l2},\dots,\theta^{ln}\} * {HN_i}(v_i;\psi_i)})
    \label{eq:new_loss}
\end{aligned}
\end{equation}
where $V = \{v_1, \dots, v_N\}$, $\Psi=\{\psi_1, \dots, \psi_N\}$.
Consequently, pFedLA transforms the optimization problem for client parameters $\theta_i$ into the hypernetwork's embedding vector $v_i$ and parameters $\psi_i$. 
In the following, we introduce the update rules of $V$ and $\Psi$.
% in the proposed pFedLA algorithm. 
% Algorithm \ref{alg:1} illustrates the proposed pFedLA algorithm.
%%%%%%%%%%%%%%%%%%%%%%%%%%%%%%%%%%%%%%%%%%%%%%%%%%%%
% \vspace{-0.3cm}
%%%%%%%%%%%%%%%%%%%%%%%%%%%%%%%%%%%%%%%%%%%%%%%%%%%%

\textbf{Update $v_i$ and $\psi_i$}. According to the chain rule, we can have the gradient of $v_i$ and $\psi_i$ from Eq.~\ref{eq:new_loss}:
\begin{equation}
\begin{aligned}
    \nabla_{v_i}\mathcal{L}_{i} = (\nabla_{v_i}\bar{\theta}_i)^T\nabla_{\bar{\theta}_i}\mathcal{L}_i \\= [\{\theta^{l1},\theta^{l2},\dots,\theta^{ln}\} * \nabla_{v_i}{HN_i}(v_i;\psi_i)]^{T}\nabla_{\bar{\theta}_i}\mathcal{L}_i,    
    \label{eq:update_of_v}
\end{aligned}
\end{equation}
\begin{equation}
\begin{aligned}
    \nabla_{\psi_i}\mathcal{L}_{i} = (\nabla_{\psi_i}\bar{\theta}_i)^T\nabla_{\bar{\theta}_i}\mathcal{L}_i \\= [\{\theta^{l1},\theta^{l2},\dots,\theta^{ln}\} * \nabla_{\psi_i}{HN_i}(v_i;\psi_i)]^{T}\nabla_{\bar{\theta}_i}\mathcal{L}_i.   
    \label{eq:update_of_psi}
\end{aligned}
\end{equation}
$\nabla_{\bar{\theta}_i}\mathcal{L}_i$ can be obtained from client $i$'s local training in each communication round and $\nabla_{v_i/\psi_i}{HN_i}(v_i;\psi_i)$ is the gradient of $\alpha_i$ in directions ${v_i/\psi_i}$. pFedLA uses a more general way to update $v_i$ and $\psi_i$:
\begin{equation}
\begin{aligned}
    \Delta{v_i} = (\nabla_{v_i}\bar{\theta}_i)^T\Delta{\theta}_i \\= [\{\theta^{l1},\theta^{l2},\dots,\theta^{ln}\} * \nabla_{v_i}{HN_i}(v_i;\psi_i)]^{T}\Delta{\theta}_i,    
    \label{eq:real_update_of_v}
\end{aligned}
\end{equation}
\begin{equation}
\begin{aligned}
    \Delta{\psi_i} = (\nabla_{\psi_i}\bar{\theta}_i)^T\Delta{\theta}_i \\= [\{\theta^{l1},\theta^{l2},\dots,\theta^{ln}\} * \nabla_{\psi_i}{HN_i}(v_i;\psi_i)]^{T}\Delta{\theta}_i.
    \label{eq:real_update_of_psi}
\end{aligned}
\end{equation}
where $\Delta{\theta}_i$ is the change of model parameters in client $i$ after local training. In accordance with Eq.~\ref{eq:real_update_of_v} and \ref{eq:real_update_of_psi}, pFedLA updates the embedding vector and parameters of hypernetwork for client $i$ at each communication round, and then update the aggregation weight matrix $\alpha_i$.

Algorithm~\ref{alg:1} demonstrates the pFedLA procedure. In each communication round, the clients first download the latest personalized models from the server, then use local SGD to train several epochs based on the private data. After that, the model update $\Delta\theta_i$ for each client will be uploaded to the server to update the embedding vector $V$ and the parameter $\Psi$. 

\begin{algorithm}[t]
% \small
\caption{HeurpFedLA Algorithm}
\label{alg:2}
\begin{algorithmic}[1]
    \Require dataset $\{D_1, D_2,\dots,D_N\}$, learning rate $\eta$. Total communication rounds $T$.
	\Ensure Trained personalized models $\{\bar{\theta}_1,\bar{\theta}_2,\dots,\bar{\theta}_N\}$.
	\State Initialize the clients' model parameters, hypernetworks parameters and embedding vectors.
% \small
\Procedure{Server executes}  {}
    \For {each communication round $t \in \{1,\dots,T\}$}
    	\For {each client $i$ \textbf{in parallel}}
    		\State $\bar{\theta}_i^{(t+1)}=\{\theta^{l1},\dots,\theta^{ln}\}*{HN_i}(v_i^{(t)};\psi_i^{(t)})$    	   
    		\State Sort $\{\alpha_i^{l1,i},\dots,\alpha_i^{ln,i}\}$ and obtain $\bar{\theta}_i^{retain}$
    		\State Set $Heur\bar{\theta}_i^{(t+1)} \leftarrow \bar{\theta}_i^{(t+1)}$ not in $\bar{\theta}_i^{retain}$
    		\State $\Delta\theta_i \leftarrow ClientUpdate(Heur\bar{\theta}_i^{(t+1)})$
        	\State Update $\{\theta^{l1},\theta^{l2},\dots,\theta^{ln}\}$ according to $\Delta\theta_i$
        	%\State $v_i^{(t+1)} = v_i^{(t)} - \eta_2[\{\theta^{l1},\theta^{l2},\dots,\theta^{ln}\} * \nabla_{v_i}{HN_i}(v_i^{(t)};\psi_i^{(t)})]^{T}\Delta\theta_i$
        	\State Update $v_i^{(t+1)}$ and $\psi_i^{(t+1)}$ via Eq.~\ref{eq:real_update_of_v}, \ref{eq:real_update_of_psi}	
        	%\State $\psi_i^{(t+1)} = \psi_i^{(t)} - \eta_2[\{\theta^{l1},\theta^{l2},\dots,\theta^{ln}\} * \nabla_{\psi_i}{HN_i}(v_i^{(t)};\psi_i^{(t)})]^{T}\Delta\theta_i$
    	\EndFor
    \EndFor
\EndProcedure

\Procedure{ClientUpdate} {$\bar{\theta}_i^{(t+1)}$}
    \State Client $i$ receives $Heur\bar{\theta}_i^{(t+1)}$ from the server.
    \State Set $\theta_i \leftarrow \{ Heur\bar{\theta}_i^{(t+1)}, \theta_i^{retain}\}$.
		\For {each local epoch }
    		\For {mini-batch $\xi_t \subseteq D_i$} 
    		  \State \textbf{Local Training:} $\theta_i = \theta_i - {\eta} \nabla_{\theta_i}\mathcal{L}_{i}(\theta_i;\xi_t)$
    		\EndFor
    	\EndFor
    \Return $\Delta\theta_i = \theta_i - \{ Heur\bar{\theta}_i^{(t+1)}, \theta_i^{retain}\}$
\EndProcedure
\end{algorithmic}
\end{algorithm}

\begin{figure}[t]
\centering
    \includegraphics[width=0.48\textwidth]{ 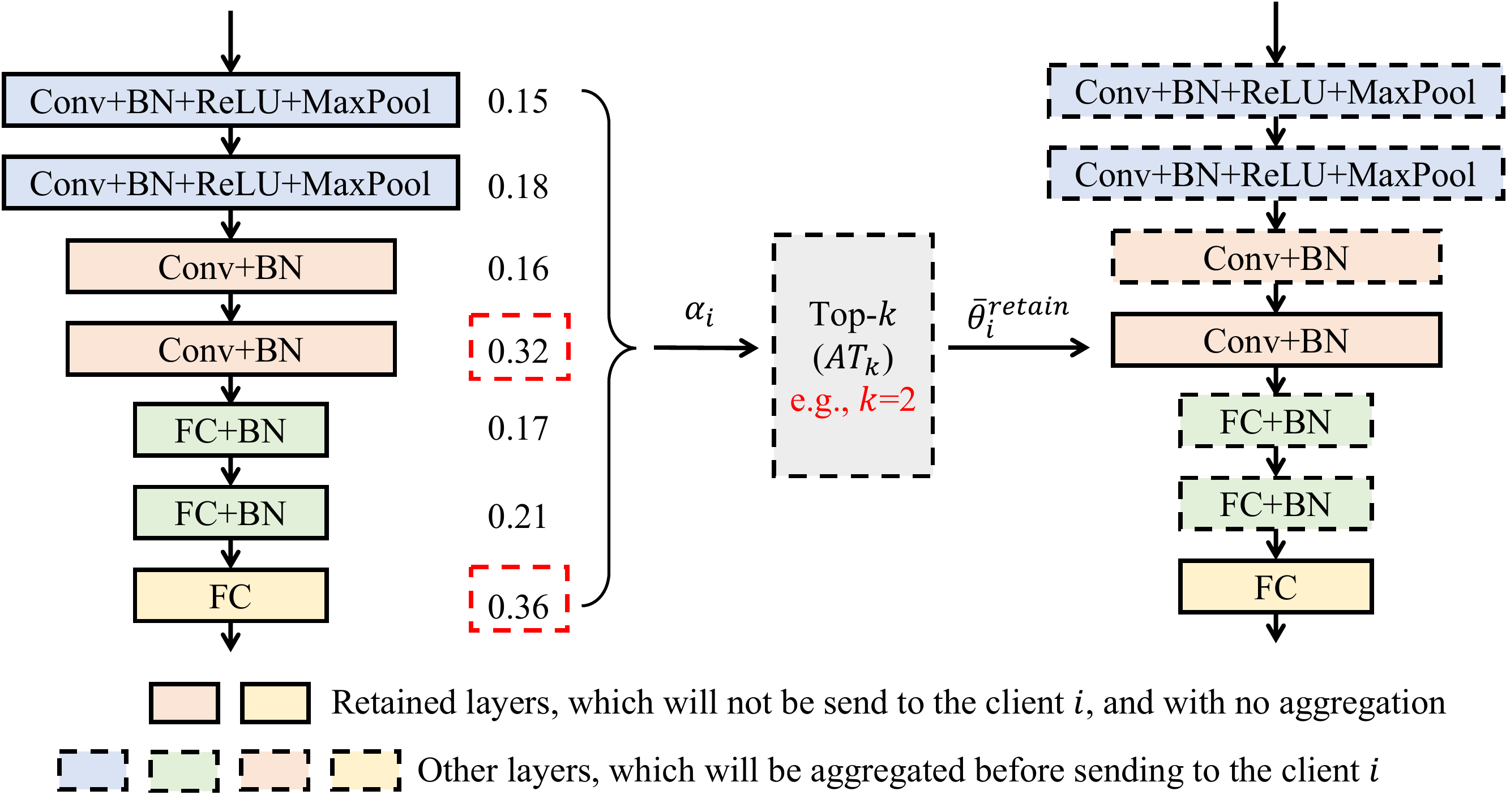}
    \caption{Illustration of top $k$ mechanism in HeurpFedLA. The selected top $k$ layers (i.e., retained layers) do not perform aggregation process, while the remaining layers execute the same operations as in pFedLA.}
    \label{fig:topk}
    \vspace{-0.3cm}
\end{figure}

\begin{table*}[t]
\centering
\small
\setlength{\tabcolsep}{1.0mm}{
\caption{Average model accuracy on 10 and 100 clients over four different datasets(non-IID\_1), respectively.}
\begin{tabular}{lcc|cc|cc|cc}
\toprule
\multicolumn{1}{c}{ }  & \multicolumn{2}{c}{EMNIST (\%)} & \multicolumn{2}{c}{FashionMNIST (\%)} &
\multicolumn{2}{c}{CIFAR10 (\%)} & \multicolumn{2}{c}{CIFAR100 (\%)}   \\ 
\cmidrule(lr){2-3}\cmidrule(lr){4-5} \cmidrule(lr){6-7}\cmidrule(lr){8-9}
 \# Clients & 10 & 100 & 10 & 100 & 10 & 100 & 10 & 100   \\ 
 \midrule
 Local Training & 89.01\textcolor[RGB]{96,96,96}{$\pm${0.47}}  & 
 91.25\textcolor[RGB]{96,96,96}{$\pm${0.18}}  & 85.83\textcolor[RGB]{96,96,96}{$\pm${0.17}}  &
 89.27\textcolor[RGB]{96,96,96}{$\pm${0.21}} & 59.44\textcolor[RGB]{96,96,96}{$\pm${0.40}}  & 64.19\textcolor[RGB]{96,96,96}{$\pm${0.19}} & 41.68\textcolor[RGB]{96,96,96}{$\pm${0.89}}  & 42.53\textcolor[RGB]{96,96,96}{$\pm${0.44}}   \\ 
 FedAvg \cite{mcmahan2017communication} & 90.45\textcolor[RGB]{96,96,96}{$\pm${0.76}}  &
 93.71\textcolor[RGB]{96,96,96}{$\pm${0.38}}  & 91.24\textcolor[RGB]{96,96,96}{$\pm${0.98}}  & 98.36\textcolor[RGB]{96,96,96}{$\pm${0.26}} & 48.57\textcolor[RGB]{96,96,96}{$\pm${0.63}}  & 58.43\textcolor[RGB]{96,96,96}{$\pm${0.29}}  & 36.64\textcolor[RGB]{96,96,96}{$\pm${0.67}}  & 
 45.19\textcolor[RGB]{96,96,96}{$\pm${0.33}}   \\
 Per-FedAvg \cite{DBLP:conf/nips/0001MO20} & 92.58\textcolor[RGB]{96,96,96}{$\pm${0.28}}  &
 92.38\textcolor[RGB]{96,96,96}{$\pm${1.14}}  & 93.63\textcolor[RGB]{96,96,96}{$\pm${1.83}}  & 92.35\textcolor[RGB]{96,96,96}{$\pm${1.55}} & 52.54\textcolor[RGB]{96,96,96}{$\pm${1.79}}  & 59.54\textcolor[RGB]{96,96,96}{$\pm${0.39}}  & 38.79\textcolor[RGB]{96,96,96}{$\pm${1.89}}  & 
 43.72\textcolor[RGB]{96,96,96}{$\pm${0.25}}   \\ 
 pFedMe \cite{t2020personalized}  &
 92.42\textcolor[RGB]{96,96,96}{$\pm${0.44}}  & 94.36\textcolor[RGB]{96,96,96}{$\pm${0.50}}  &
 90.43\textcolor[RGB]{96,96,96}{$\pm${0.86}}  & 98.57\textcolor[RGB]{96,96,96}{$\pm${0.38}}& 53.73\textcolor[RGB]{96,96,96}{$\pm${3.74}}  & 65.97\textcolor[RGB]{96,96,96}{$\pm${1.61}}  & 42.29\textcolor[RGB]{96,96,96}{$\pm${3.67}}  & 53.60\textcolor[RGB]{96,96,96}{$\pm${1.28}}   \\ 
 pFedHN \cite{DBLP:conf/icml/ShamsianNFC21} & 93.94\textcolor[RGB]{96,96,96}{$\pm${0.16}}  &
 \textbf{96.64}\textcolor[RGB]{96,96,96}{$\pm${0.91}}  & 94.83\textcolor[RGB]{96,96,96}{$\pm${0.33}}  & 98.80\textcolor[RGB]{96,96,96}{$\pm${0.92}}& 46.98\textcolor[RGB]{96,96,96}{$\pm${1.91}}  & 63.71\textcolor[RGB]{96,96,96}{$\pm${1.26}}  & 39.67\textcolor[RGB]{96,96,96}{$\pm${0.52}}  & 
 51.36\textcolor[RGB]{96,96,96}{$\pm${1.77}}    \\ 
 FedBN \cite{DBLP:conf/iclr/LiJZKD21} & -  &
 -  & -  & -& 59.36\textcolor[RGB]{96,96,96}{$\pm${0.92}}  & 70.88\textcolor[RGB]{96,96,96}{$\pm${0.36}}  & 45.18\textcolor[RGB]{96,96,96}{$\pm${0.42}}  & 
 56.16\textcolor[RGB]{96,96,96}{$\pm${0.38}}    \\ 
 FedRep \cite{DBLP:conf/icml/CollinsHMS21} & 91.82\textcolor[RGB]{96,96,96}{$\pm${0.15}}  &
 95.23\textcolor[RGB]{96,96,96}{$\pm${0.12}}  & 93.17\textcolor[RGB]{96,96,96}{$\pm${0.26}}  & 97.15\textcolor[RGB]{96,96,96}{$\pm${0.09}} & 58.01\textcolor[RGB]{96,96,96}{$\pm${0.56}}  & 71.94\textcolor[RGB]{96,96,96}{$\pm${0.22}}  & 44.33\textcolor[RGB]{96,96,96}{$\pm${0.63}}  & 
 56.47\textcolor[RGB]{96,96,96}{$\pm${0.41}}   \\ 
 FedFomo \cite{zhang2020personalized} & 88.33\textcolor[RGB]{96,96,96}{$\pm${0.29}}  &
 91.36\textcolor[RGB]{96,96,96}{$\pm${0.17}}  & 86.17\textcolor[RGB]{96,96,96}{$\pm${0.34}}  & 91.83\textcolor[RGB]{96,96,96}{$\pm${0.12}} & 59.37\textcolor[RGB]{96,96,96}{$\pm${0.71}}  & 66.07\textcolor[RGB]{96,96,96}{$\pm${0.24}}  & 41.89\textcolor[RGB]{96,96,96}{$\pm${0.78}}  & 
 44.28\textcolor[RGB]{96,96,96}{$\pm${0.28}}   \\ 
%  \cline{1-9}
\midrule
  pFedLA (Ours) & {90.65}\textcolor[RGB]{96,96,96}{$\pm${0.41}}  &
 96.34\textcolor[RGB]{96,96,96}{$\pm${1.35}}  & {94.34}\textcolor[RGB]{96,96,96}{$\pm${0.29}}  &
 \textbf{98.87}\textcolor[RGB]{96,96,96}{$\pm${0.66}} & \textbf{61.43}\textcolor[RGB]{96,96,96}{$\pm${0.56}}  & \textbf{73.15}\textcolor[RGB]{96,96,96}{$\pm${0.83}}  & \textbf{47.22}\textcolor[RGB]{96,96,96}{$\pm${0.77}}  & \textbf{56.62}\textcolor[RGB]{96,96,96}{$\pm${0.81}}   \\ 
   HeurpFedLA (Ours)  & \textbf{94.11}\textcolor[RGB]{96,96,96}{$\pm${0.13}}  &
 95.04\textcolor[RGB]{96,96,96}{$\pm${0.41}}  & \textbf{95.47}\textcolor[RGB]{96,96,96}{$\pm${0.47}}  &
 {96.95}\textcolor[RGB]{96,96,96}{$\pm${0.44}}& 60.02\textcolor[RGB]{96,96,96}{$\pm${0.74}}  & 73.05\textcolor[RGB]{96,96,96}{$\pm${1.02}}  & 46.47\textcolor[RGB]{96,96,96}{$\pm${0.83}}  & 54.43\textcolor[RGB]{96,96,96}{$\pm${1.37}}   \\ 
\bottomrule
 \label{tab:1}
 \vspace{-0.3cm}
\end{tabular}}
\end{table*}

\begin{table*}[t]
\centering
\small
\setlength{\tabcolsep}{1.0mm}{
\caption{Average model accuracy on 10 and 100 clients over four different datasets(non-IID\_2), respectively.}
\begin{tabular}{lcc|cc|cc|cc}
\toprule
\multicolumn{1}{c}{ } & \multicolumn{2}{c}{EMNIST (\%)} & \multicolumn{2}{c}{FashionMNIST (\%)} & 
\multicolumn{2}{c}{CIFAR10 (\%)} & \multicolumn{2}{c}{CIFAR100 (\%)}  \\ 
\cmidrule{2-9}
 \# Clients & 10 & 100 & 10 & 100 & 10 & 100 & 10 & 100   \\ 
\midrule
 Local Training  & 80.72\textcolor[RGB]{96,96,96}{$\pm${0.43}}  & 
 79.09\textcolor[RGB]{96,96,96}{$\pm${0.12}}  & 65.60\textcolor[RGB]{96,96,96}{$\pm${0.59}}  &
 65.97\textcolor[RGB]{96,96,96}{$\pm${0.28}}& 39.79\textcolor[RGB]{96,96,96}{$\pm${0.42}}  & 45.15\textcolor[RGB]{96,96,96}{$\pm${0.29}} & 26.29\textcolor[RGB]{96,96,96}{$\pm${0.37}}  & 27.87\textcolor[RGB]{96,96,96}{$\pm${0.28}}   \\ 
 FedAvg \cite{mcmahan2017communication} & 90.43\textcolor[RGB]{96,96,96}{$\pm${0.58}}  &
 93.91\textcolor[RGB]{96,96,96}{$\pm${0.32}}  & 89.09\textcolor[RGB]{96,96,96}{$\pm${0.57}}  & 98.25\textcolor[RGB]{96,96,96}{$\pm${0.38}}& 44.89\textcolor[RGB]{96,96,96}{$\pm${0.21}}  & 54.03\textcolor[RGB]{96,96,96}{$\pm${0.37}}  & 32.24\textcolor[RGB]{96,96,96}{$\pm${0.74}}  & 
 40.89\textcolor[RGB]{96,96,96}{$\pm${0.46}}    \\
 Per-FedAvg \cite{DBLP:conf/nips/0001MO20}& 90.86\textcolor[RGB]{96,96,96}{$\pm${0.78}}  &
 94.09\textcolor[RGB]{96,96,96}{$\pm${0.18}}  & 90.78\textcolor[RGB]{96,96,96}{$\pm${1.12}}  & 98.53\textcolor[RGB]{96,96,96}{$\pm${0.95}}  & 44.48\textcolor[RGB]{96,96,96}{$\pm${0.82}}  & 54.40\textcolor[RGB]{96,96,96}{$\pm${0.44}}  & 30.86\textcolor[RGB]{96,96,96}{$\pm${1.11}}  & 
 42.56\textcolor[RGB]{96,96,96}{$\pm${0.28}}   \\ 
 pFedMe \cite{t2020personalized} & 89.13\textcolor[RGB]{96,96,96}{$\pm${0.58}}  & 93.87\textcolor[RGB]{96,96,96}{$\pm${0.40}}  &
 85.15\textcolor[RGB]{96,96,96}{$\pm${0.94}}  & 97.87\textcolor[RGB]{96,96,96}{$\pm${0.19}}& 46.97\textcolor[RGB]{96,96,96}{$\pm${1.19}}  & 58.23\textcolor[RGB]{96,96,96}{$\pm${1.07}}  & 33.45\textcolor[RGB]{96,96,96}{$\pm${0.86}}  & 44.35\textcolor[RGB]{96,96,96}{$\pm${0.96}}  
  \\ 
 pFedHN \cite{DBLP:conf/icml/ShamsianNFC21} & 91.37\textcolor[RGB]{96,96,96}{$\pm${0.41}}  &
 94.48\textcolor[RGB]{96,96,96}{$\pm${0.51}}  & 93.45\textcolor[RGB]{96,96,96}{$\pm${0.11}}  & \textbf{98.83}\textcolor[RGB]{96,96,96}{$\pm${0.82}} & 37.49\textcolor[RGB]{96,96,96}{$\pm${0.94}}  & 49.90\textcolor[RGB]{96,96,96}{$\pm${1.66}}  & 26.35\textcolor[RGB]{96,96,96}{$\pm${0.93}}  & 
 40.27\textcolor[RGB]{96,96,96}{$\pm${0.82}}   \\ 
 FedBN \cite{DBLP:conf/iclr/LiJZKD21}& -  &
 -  & -  & -  & 49.79\textcolor[RGB]{96,96,96}{$\pm${0.33}}  & 60.62\textcolor[RGB]{96,96,96}{$\pm${0.42}}  & 34.94\textcolor[RGB]{96,96,96}{$\pm${0.50}}  & 
 46.42\textcolor[RGB]{96,96,96}{$\pm${0.54}}   \\ 
 FedRep \cite{DBLP:conf/icml/CollinsHMS21} & 86.81\textcolor[RGB]{96,96,96}{$\pm${0.29}}  &
 90.32\textcolor[RGB]{96,96,96}{$\pm${0.08}}  & 79.13\textcolor[RGB]{96,96,96}{$\pm${0.56}}  & 92.04\textcolor[RGB]{96,96,96}{$\pm${0.23}}& 49.16\textcolor[RGB]{96,96,96}{$\pm${0.73}}  & 60.36\textcolor[RGB]{96,96,96}{$\pm${0.57}}  & 34.19\textcolor[RGB]{96,96,96}{$\pm${0.74}}  & 
 43.51\textcolor[RGB]{96,96,96}{$\pm${0.34}}    \\ 
 FedFomo \cite{zhang2020personalized} & 80.14\textcolor[RGB]{96,96,96}{$\pm${0.42}}  &
 82.61\textcolor[RGB]{96,96,96}{$\pm${0.11}}  & 64.10\textcolor[RGB]{96,96,96}{$\pm${0.38}}  & 67.91\textcolor[RGB]{96,96,96}{$\pm${0.29}} & 40.62\textcolor[RGB]{96,96,96}{$\pm${0.31}}  & 47.08\textcolor[RGB]{96,96,96}{$\pm${0.49}}  & 27.33\textcolor[RGB]{96,96,96}{$\pm${0.51}}  & 
 29.63\textcolor[RGB]{96,96,96}{$\pm${0.24}}   \\ 
\midrule
  pFedLA (Ours) & \textbf{92.06}\textcolor[RGB]{96,96,96}{$\pm${0.71}}  &
 \textbf{94.83}\textcolor[RGB]{96,96,96}{$\pm${1.04}}  & \textbf{93.89}\textcolor[RGB]{96,96,96}{$\pm${0.91}}  &
 98.41\textcolor[RGB]{96,96,96}{$\pm${0.98}} & \textbf{49.93}\textcolor[RGB]{96,96,96}{$\pm${0.96}}  & \textbf{61.82}\textcolor[RGB]{96,96,96}{$\pm${1.89}}  & {35.02}\textcolor[RGB]{96,96,96}{$\pm${0.83}}  & \textbf{48.79}\textcolor[RGB]{96,96,96}{$\pm${1.60}}   \\ 
   HeurpFedLA (Ours) & {91.98}\textcolor[RGB]{96,96,96}{$\pm${0.36}}  &
 93.31\textcolor[RGB]{96,96,96}{$\pm${0.77}}  & {92.01}\textcolor[RGB]{96,96,96}{$\pm${0.74}}  &
 {98.66}\textcolor[RGB]{96,96,96}{$\pm${0.80}}& 49.06\textcolor[RGB]{96,96,96}{$\pm${0.68}}  & 60.62\textcolor[RGB]{96,96,96}{$\pm${1.73}}  & \textbf{35.42}\textcolor[RGB]{96,96,96}{$\pm${0.49}}  & 48.72\textcolor[RGB]{96,96,96}{$\pm${1.75}}    \\ 
\bottomrule
 \label{tab:2}
 \vspace{-0.3cm}
\end{tabular}}
\end{table*}

\subsection{HeurpFedLA: Heuristic Improvement of pFedLA on Communication Efficiency}
The communication overhead of pFedLA is determined by the size of $\Delta\theta_i$ sent from the clients and $\bar{\theta}_i$ sent from the server. So, there is no additional communication cost comparing with traditional FL methods, e.g., FedAvg. In this section, we propose to further reduce the communication overhead of pFedLA with negligible performance reduction, which can adapt to more general scenarios, e.g., large scale FL systems, limited communication capacities, etc. 
% However,
% in large-scale FL systems, limited communication bandwidth,  and unreliable device availability are still challenges and need to be tackled. 
%
% To improve the scalability in large-scale FL system, in this section, we propose a heuristic improvement of pFedLA to achieve communication efficiency.
% Since hypernetworks are located on the server, pFedLA does not cause additional data transfer compared to FedAvg.

%JIe  (the whole subsection)
Comparing with existing works that keep some specific layers updated locally to enable communication-efficient training while retaining the performance of pFL \cite{DBLP:conf/icml/CollinsHMS21,DBLP:conf/iclr/LiJZKD21,liang2020think}, e.g., FedBN \cite{DBLP:conf/iclr/LiJZKD21} found that local models with BN layers should exclude these parameters from the aggregating steps during training, while FedRep \cite{DBLP:conf/icml/CollinsHMS21} and LG-FedAvg \cite{liang2020think} proposed to locally learn the classifier layer and representation layers respectively, pFedLA can give an alternative guidance to determine which layers should be retained locally. 
To this end, we propose HeurpFedLA, a heuristic improvement of pFedLA that partial layers are retained locally, and the remaining layers are aggregated at the server side during training.
The key idea of HeurpFedLA is to heuristically select the partial layers $\bar{\theta}_i^{retain}$ with top $k$ ($\text{AT}_k$) aggregation weights to update locally. Specifically, by using the aggregation weights $\alpha_i^{l1,i}, \alpha_i^{l2,i}, \dots, \alpha_i^{ln,i}$ for all layers of client $i$, we can sort these weights in descending order and select corresponding top $k$ layers
\begin{equation}
    \bar{\theta}_i^{retain} = \text{AT}_k\{\bar{\theta}_i^{l1}, \dots, \bar{\theta}_i^{ln}|\alpha_i^{l1,i}, \dots, \alpha_i^{ln,i}\},
    \label{eq:top_k}
\end{equation}
where $\text{AT}_k$ is the top $k$ selection function described above, and $k$ is a hyperparameter manually denoted before training. The detailed workflow of top $k$ selection mechanism is shown in Figure~\ref{fig:topk}.

The principle behind HeurpFedLA is that layers with higher rank index should contribute more to the model personalization, which means directly using these layers in personalized model has little impact on the training performance.
% Applying top $k$ strategy by HeurpFedLA brings several benefits: 1) since   
%
The retention of local layers by HeurpFedLA brings benefits in terms of communication overhead reduction from the server to the clients direction, i.e., the server can save the costs of transmitting the parameters of the retained layers.

% . Although it has no effect on the upward communication from client to server, in the downward communication from server to client, the server does not need to send the parameters of the retained layers. When the number of parameters in retained layers is large, the huge number of communication overhead can be saved. 

As to be demonstrated in Section \ref{subsec:communication}, HeurpFedLA can significantly reduce the communication cost while maintaining the model performance of pFL. In large scale FL systems, it is of practical value to keep some layers from aggregation and transmission, especially for limited communication bandwidth scenarios.   
Furthermore, HeurpFedLA  is a general training framework and can be effectively compatible with common compression schemes such as gradient quantization, sparsification, etc.
The impact of retaining local layers is discussed in more detail in the next section.

\section{Evaluation}
\label{sec:evaluation}

\begin{figure*}[t]
  \centering
  \begin{subfigure}{0.24\linewidth}
    % \fbox{\rule{0pt}{2in} \rule{.9\linewidth}{0pt}}
    \centering
    \includegraphics[width = 0.95\linewidth]{ 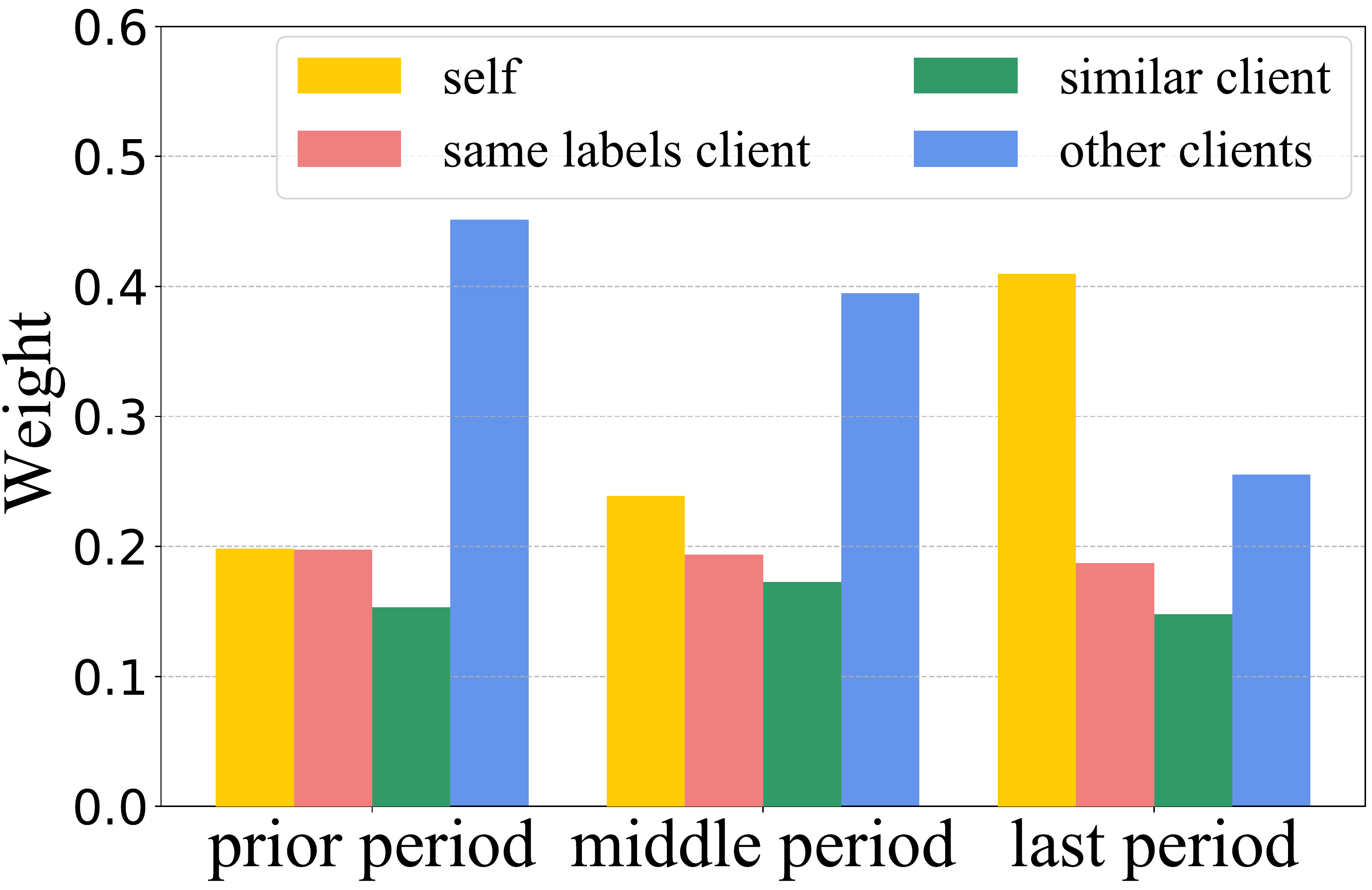}
    \caption{EMNIST}
    \label{fig:short-a}
  \end{subfigure}
  \hfill
  \begin{subfigure}{0.24\linewidth}
    % \fbox{\rule{0pt}{2in} \rule{.9\linewidth}{0pt}}
    \centering
    \includegraphics[width = 0.95\linewidth]{ 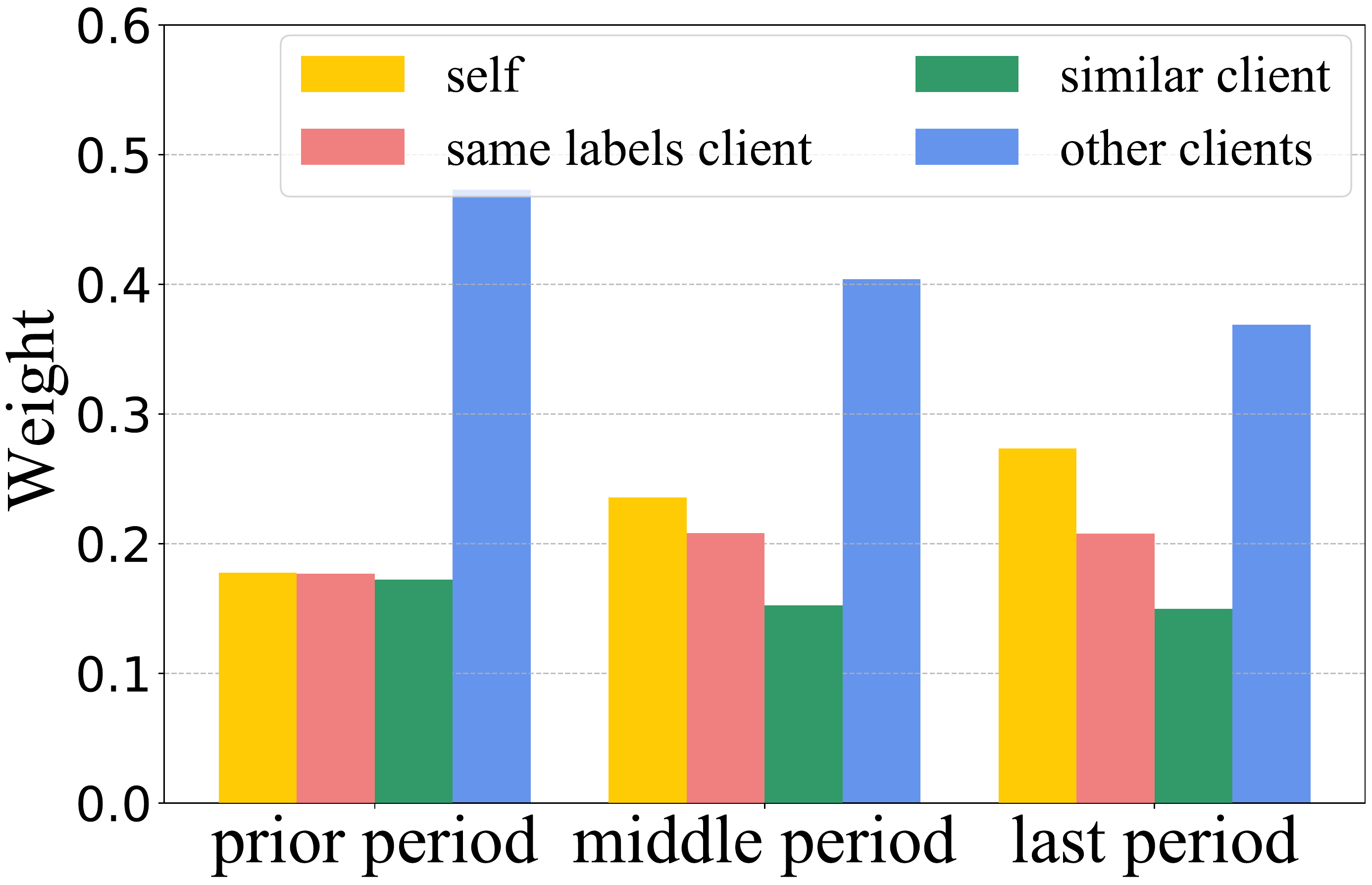}
    \caption{FashionMNIST}
    \label{fig:short-b}
  \end{subfigure}
  \hfill
  \begin{subfigure}{0.24\linewidth}
    % \fbox{\rule{0pt}{2in} \rule{.9\linewidth}{0pt}}
    \centering
    \includegraphics[width = 0.95\linewidth]{ 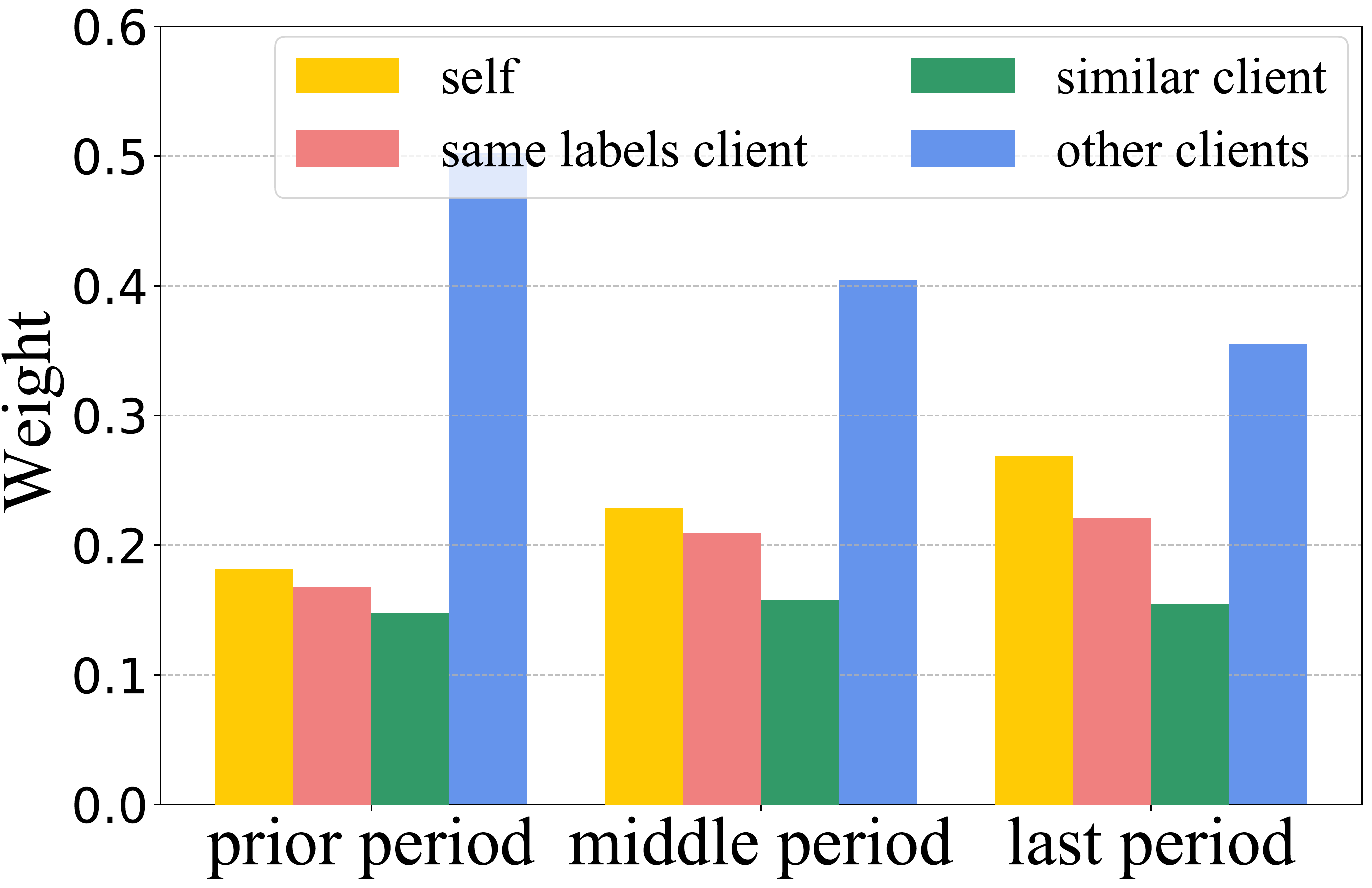}
    \caption{CIFAR10}
    \label{fig:short-c}
  \end{subfigure}
  \hfill
  \begin{subfigure}{0.24\linewidth}
    % \fbox{\rule{0pt}{2in} \rule{.9\linewidth}{0pt}}
    \centering
    \includegraphics[width = 0.95\linewidth]{ 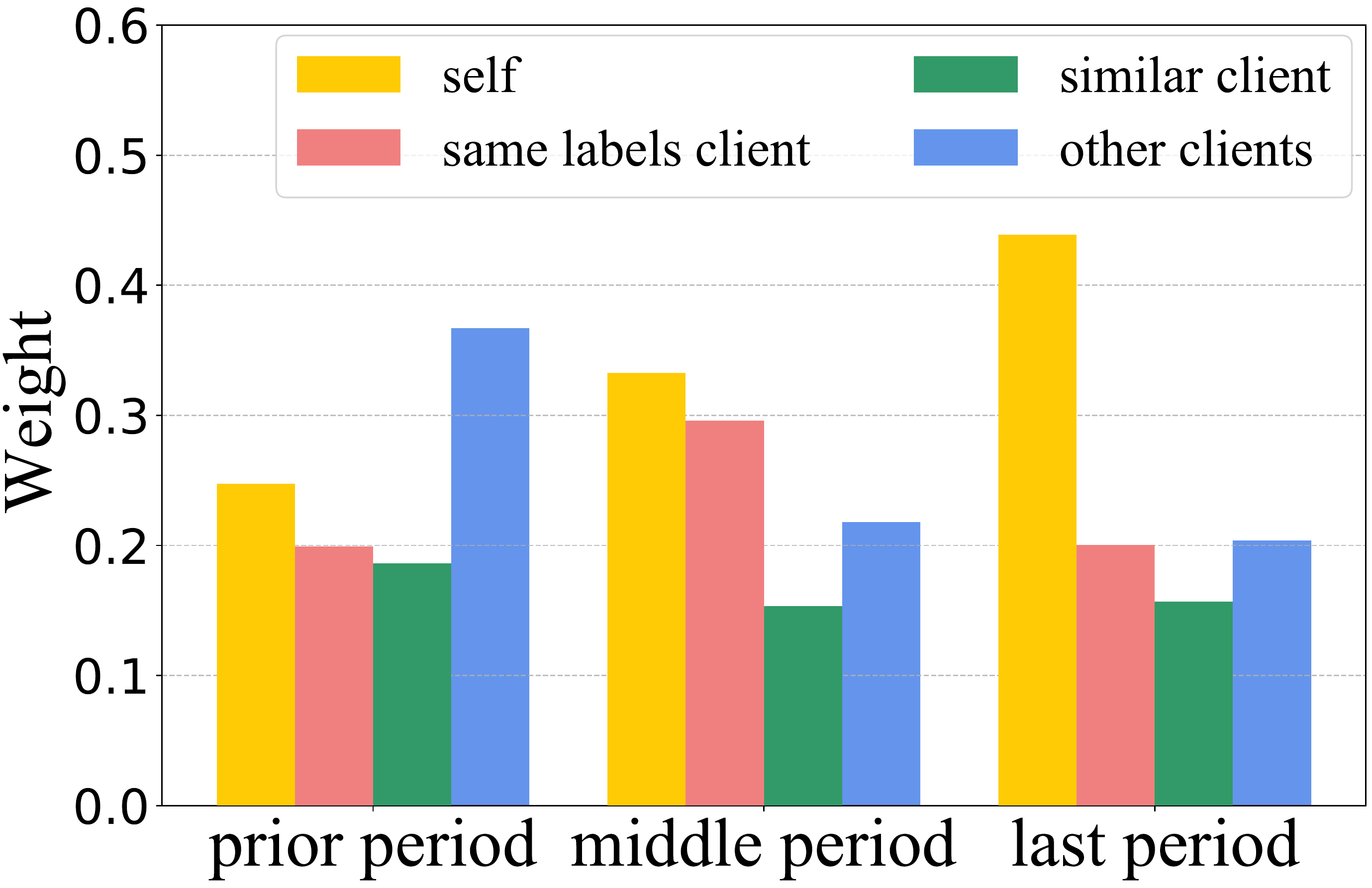}
    \caption{CIFAR100}
    \label{fig:short-d}
  \end{subfigure}
  \caption{Change of aggregation weights during the prior, middle and last period of training phase.}
  \label{fig:weight}
    \vspace{-0.3cm}
\end{figure*}

\begin{figure*}[t]
  \centering
  \begin{subfigure}[b]{0.24\linewidth}
    % \fbox{\rule{0pt}{2in} \rule{.9\linewidth}{0pt}}
    \centering
    \includegraphics[width = 0.95\linewidth]{ 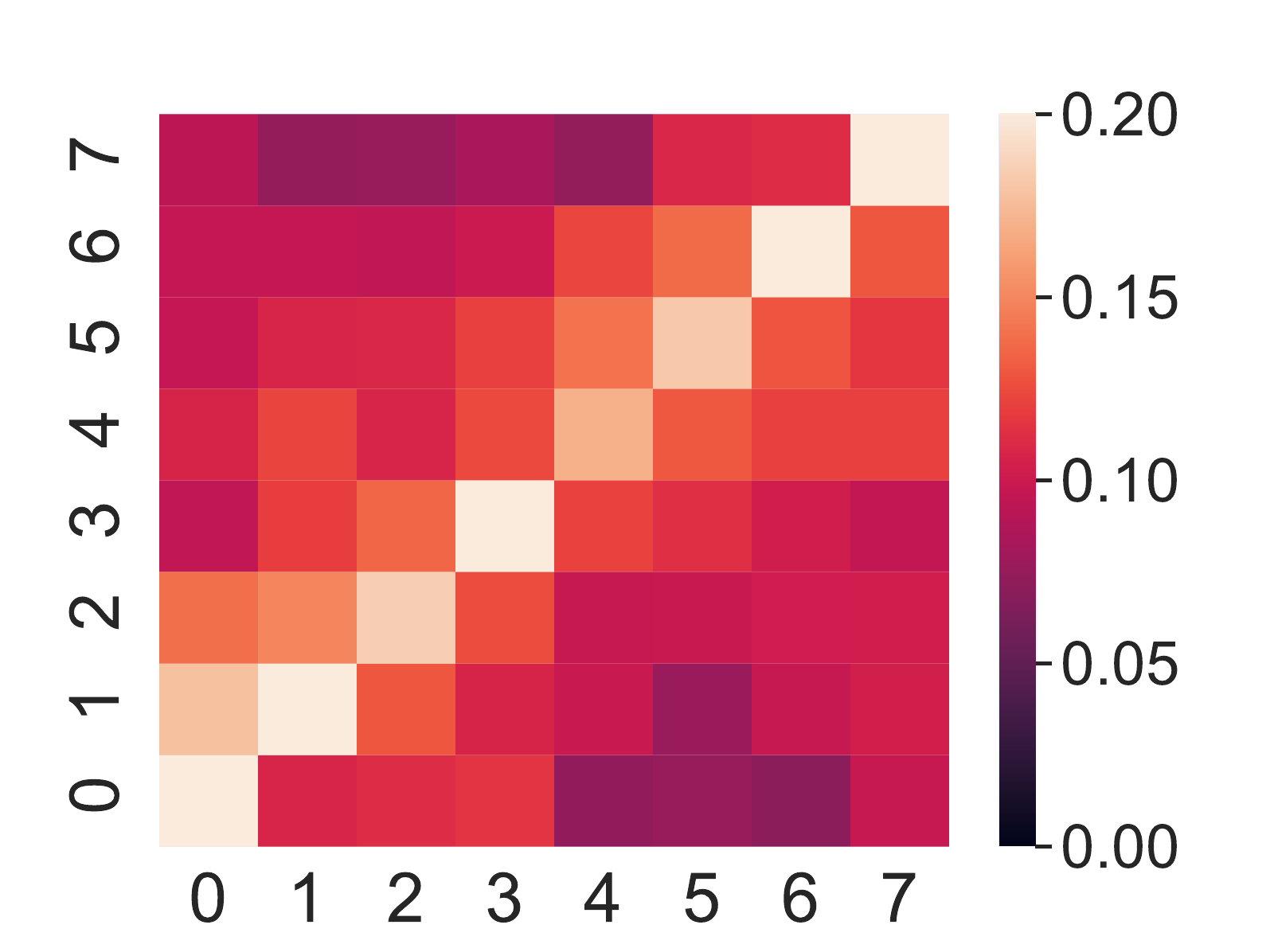}
    \caption{EMNIST (FC1)}
    \label{fig:heatmap-a}
  \end{subfigure}
  \hfill
  \begin{subfigure}[b]{0.24\linewidth}
    % \fbox{\rule{0pt}{2in} \rule{.9\linewidth}{0pt}}
    \centering
    \includegraphics[width = 0.95\linewidth]{ 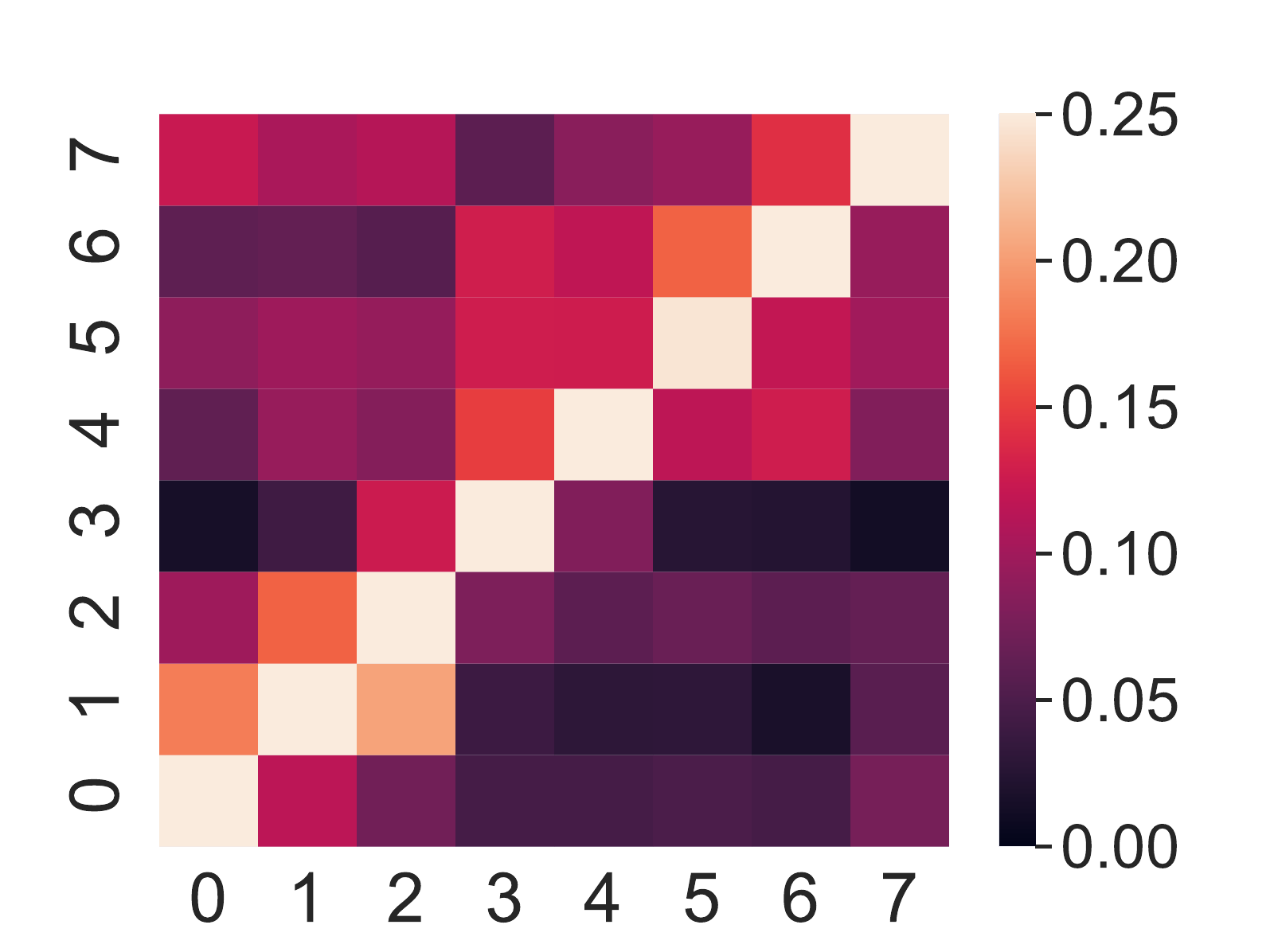}
    \caption{FashionMNIST (FC1)}
    \label{fig:heatmap-b}
  \end{subfigure}
  \hfill
  \begin{subfigure}[b]{0.24\linewidth}
    % \fbox{\rule{0pt}{2in} \rule{.9\linewidth}{0pt}}
    \centering
    \includegraphics[width = 0.95\linewidth]{ 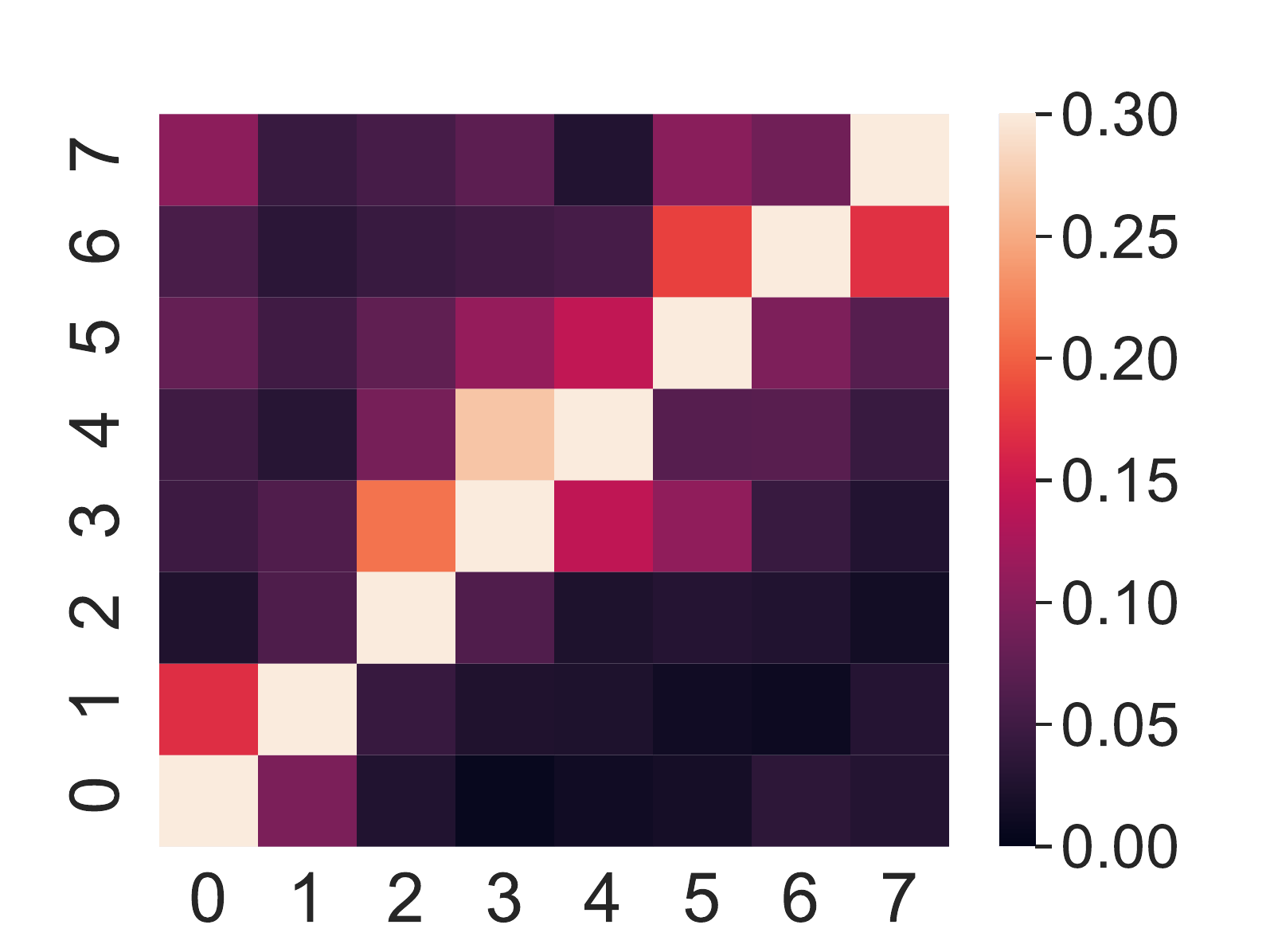}
    \caption{CIFAR10 (FC3)}
    \label{fig:heatmap-c}
  \end{subfigure}
  \hfill
  \begin{subfigure}[b]{0.24\linewidth}
    % \fbox{\rule{0pt}{2in} \rule{.9\linewidth}{0pt}}
    \centering
    \includegraphics[width = 0.95\linewidth]{ 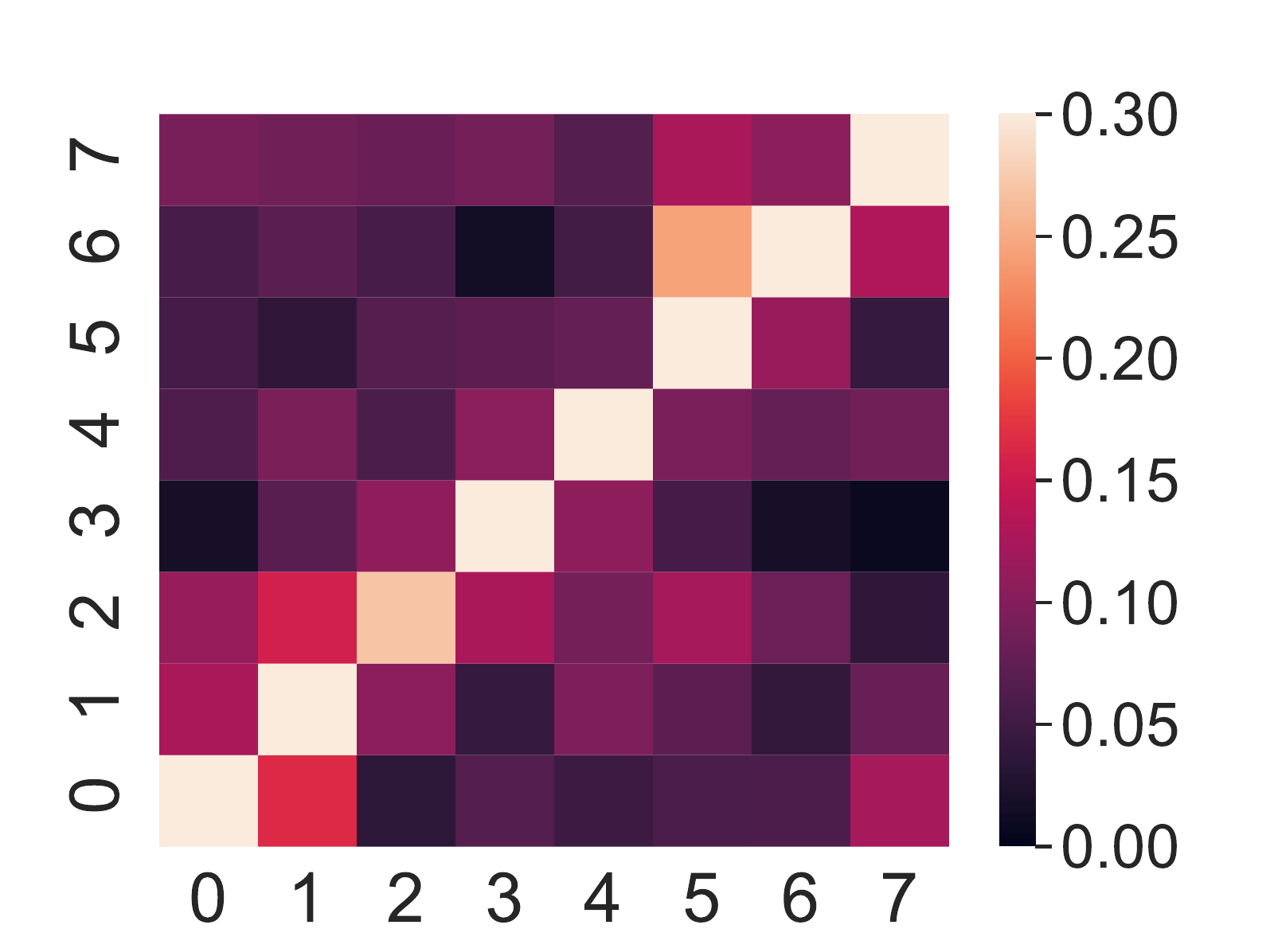}
    \caption{CIFAR100 (FC3)}
    \label{fig:heatmap-d}  
  \end{subfigure}
  \caption{The visualization of the aggregation weights in a specific layer on EMNIST, FashionMNIST, CIFAR10 and CIFAR100. X-axis and y-axis show the IDs of clients.}
  \label{fig:weight_visualization}
  \vspace{-0.3cm}
\end{figure*}

\subsection{Experimental Setup}
\noindent\textbf{Datasets.} We evaluate the pFedLA framework over four datasets, EMNIST, FashionMNIST, CIFAR10 and CIFAR100. The distribution of all data sets on the training clients is non-IID. We consider two non-IID scenarios: 1) each client is randomly assigned four classes (twelve classes per client in CIFAR100) with the same amount of data on each class; 2) each client contains all classes, while the data on each class is not uniformly distributed. Two classes in EMNIST, FashionMNIST, CIFAR10 datasets have higher number of data samples than other classes, while six classes in CIFAR100 have more data samples than the others. All data are divided into $70\%$ training set, and $30\%$ test set. The test set and the training set have the same data distribution for all clients.

\noindent\textbf{Baselines.} We compared the performance of pFedLA and HeurpFedLA with the state-of-the-art methods. In addition to \textbf{FedAvg} and \textbf{Local Training}, we also include \textbf{Per-Fedavg}, a pFL algorithm based on meta-learning; \textbf{pFedMe}, a pFL algorithm with regularization term added in the objective function; \textbf{pFedHN}, a pFL algorithm that uses hypernetworks to directly produce personalized model; \textbf{FedBN}, keeps each client's BN layer updating locally, while other layers are aggregated according to the FedAvg algorithm; \textbf{FedRep}, a pFL algorithm that keeps each client's classifier updating locally, while the other parts are aggregated at the server; \textbf{FedFomo}, a pFL algorithm that uses distance to calculate the aggregation weights based on the model and loss differences. 

% Except for the distance between parameters, it also uses other clients’ parameters and local client's data to calculate the loss value, then evaluates other clients’ value for the local client.

\noindent\textbf{Training Details.} In all experiments, we use the same CNN architectures as in FedFomo\cite{zhang2020personalized}, FedBN\cite{DBLP:conf/iclr/LiJZKD21} and pFedHN\cite{DBLP:conf/icml/ShamsianNFC21}. All the models have the same structure between different clients under the same setting. For CIFAR10 and CIFAR100, we add BN layers after the convolutional layers. For EMNIST and FashionMNIST, there is no BN layers in the model. The hypernetwork for computing layer-wise aggregation weights is a simple structure of several fully connected layers. The weight of each layer for a target client is calculated by a corresponding fully connected layer in the hypernetwork. For the specific structure of hypernetwork, please refer to the supplemental material. 
We evaluate the performance of pFedLA in two settings, i.e., 10 clients with 100\% participation and 100 clients with 10\% participation. The average model accuracy of all clients is obtained after 600 rounds training for 10 clients case and 2500 rounds for 100 clients. 

% We evaluate under both setups with two FL scenarios: 10 and 100 clients with 100\% and 10\% participation, respectively, reporting the final accuracy after training for 600 communication rounds in the former and 2500 rounds in the latter.
% In the experiment of $10$ clients, we performed about $600$ communication rounds with a participation rate of $100\%$, while in the experiment with $100$ clients, we performed about $2500$ communication rounds with a participation rate of $10\%$.

\noindent\textbf{Implementation.} We simulate all clients and the server on a workstation with an RTX 2080Ti GPU, a $3.6$-GHZ Intel Core i9-9900KF CPU and $64$GB of RAM. All methods are implemented in PyTorch.

% \begin{figure*}[h]
% \centering
% \subcaptionbox{CIFAR10\label{1}}{\includegraphics[width = 0.35\linewidth]{ pics/py_exp2_cifar10_cj.pdf}}
% \subcaptionbox{EMNIST\label{2}}{\includegraphics[width = 0.35\linewidth]{ pics/py_exp2_emnist_cj.pdf}}
% \caption{Change of aggregation weights as training progresses }
% \label{fig:weight}
% \vspace{-0.3cm}
% \end{figure*}

% \begin{figure}[h]
% \centering
% \subcaptionbox{CIFAR10 (fc3)\label{1}}{\includegraphics[width = 0.49\linewidth]{ pics/heatmap_cifar10.pdf}}
% \subcaptionbox{EMNIST (fc1)\label{2}}{\includegraphics[width = 0.49\linewidth]{ pics/heatmap_emnist.pdf}}
% \caption{The visualization of the aggregation weights in one layer on CIFAR10 and EMNIST. X-axis and y-axis show the IDs of clients.}
% \label{fig:weight_visualization}
% \vspace{-0.3cm}
% \end{figure}

% \begin{figure*}
%   \centering
%   \begin{subfigure}{0.49\linewidth}
%   \centering
%     % \fbox{\rule{0pt}{2in} \rule{.9\linewidth}{0pt}}
%     \includegraphics[width = 0.95\linewidth]{ pics/exp2_emnist.png}
%     \caption{}
%     \label{fig:short-a}
%   \end{subfigure}
%   \begin{subfigure}{0.49\linewidth}
%   \centering
%     % \fbox{\rule{0pt}{2in} \rule{.9\linewidth}{0pt}}
%     \includegraphics[width = 0.95\linewidth]{ pics/exp2_cifar10.png}
%     \caption{}
%     \label{fig:short-b}
%   \end{subfigure}
%   \caption{Change of aggregation weights as training progresses}
%   \label{fig:short}
% \end{figure*}

\subsection{Performance Evaluation}\label{section:performance}
% \noindent\textbf{Performance Comparison.} 
%%%% Jie: need explain the experimental setting, e.g., number of epoch, learning rate\dots
For all experiments, we use cross-entropy loss and SGD optimizer with a batch size of $32$. The number of local epochs is $10$ for $10$ clients case and $20$ for $100$ clients. The learning rate is $0.01$ for CIFAR10 and CIFAR100, and $0.005$ for EMNIST and FashionMNIST. 
The performance of both the baselines and the proposed pFedLA under two different non-IID cases are listed in Table~\ref{tab:1} and Table \ref{tab:2}, respectively. Our proposed algorithm provides superior performance than baselines over the four datasets with different data distributions in most cases. On the other hand, HeurpFedLA also outperforms the existing methods with negligible performance reduction comparing with pFedLA. 
The number of retained layers ($k$) in Table~\ref{tab:1} and \ref{tab:2} is 1.
The communication costs of HeurpFedLA is discussed in Section~\ref{subsec:communication}.   
% competitive accuracy for all the four datasets with different data distributions. Other better-performing pFL algorithm cannot maintain the best accuracy in all situations. i.e., FedFomo and FedRep in Non-IID\_2; FedBn in Non-IID\_1 and pFedBN in CIFAR10 and CIFAR100. Our method, on the other hand, has good adaptability. 
%
% HeurpFedLA performs slightly less well than pFedLA in most cases, but HeurpFedLA can save considerable communication cost, as discussed in more detail in Section~\ref{subsec:communication}. HeurpFedLA is a more flexible option when there is a trade-off between performance and communication overhead. 
%
Note that since all clients have the same amount of training data for both 10 and 100 clients cases, so the 100 clients case has much more data, and thus can provide better model accuracy.

% It should be noticed that the number of data on each client in the case of 100 clients is the same as in only 10 clients. In this case, the total amount of data is increased, and thus lead to the increase of the final test accuracy.

\begin{table*}[t]
\centering
% \small
\setlength{\tabcolsep}{1.4mm}{
\caption{Average Model Accuracy and Communication Cost on different number of retained layers (i.e., $k$) over EMNIST and CIFAR10.}
\begin{tabular}{lccc|cccc}
\toprule
\multicolumn{1}{c}{ }  & 
\multicolumn{3}{c}{EMNIST} &  
\multicolumn{4}{c}{CIFAR10}  \\ 
% \cline{2-5} \cline{6-8}
\cmidrule(lr){2-4}\cmidrule(lr){5-8}
 \# Number of retained layers ($k$) & 0 & 1 & 2
 & 0 & 1 & 2 & 3\\ 
%  \ \ retained & & & & & & &\\
 \midrule
Model Accuracy (\%)  & 90.65  & 
 94.11  & 93.94 & 61.43  & 60.02 & 59.90  & 59.23  \\ 
Communication Cost (MBytes)  & 491.08  &
 488.65  & 312.52 &693.98  & 418.97  & 382.25  & 379.82   \\
\bottomrule
\label{tab:heurpfedla}
 \vspace{-0.3cm}
\end{tabular}}
\end{table*}

% \subsection{Ablation Experiments}\label{subsec:ablation}
\subsection{Analysis of Weight Evolution} 
To demonstrate that our method can generate higher weights to those clients with similar data distribution, we conduct the experiments with 8 clients who randomly 4 data classes from the corresponding datasets. From the 8 clients, we consider a target client with 4 random data classes, one contrastive client who has the same four classes, and one similar client who has 3 same classes with the target client. We record the weight value of each layer on the target client during the training process. 
%Figure~\ref{fig:weight} shows the evolution of the aggregation weights for the target client during the prior, middle and last periods of training phase. It can be observed that self-weights from the target client himself, the inter-weights from the contrastive client and the inter-weights from the similar clients all increase with the training process, while the inter-weights from other clients decrease. Besides, the inter-weights from the contrastive client are larger than that of the similar client, which shows that the hypernetwork can learn the inter-similarity of all clients in layer granularity.    
Figure~\ref{fig:weight} shows the evolution of the aggregation weights for the target client during the prior, middle and last periods of training phase. It can be observed that the inter-weights from other clients decrease with the training process because their data distribution is very different from the target client. Besides, for the target client, clients with more similar data distribution (e.g., the same labels client) have higher weight value than other clients (e.g., the similar client), which shows that the hypernetwork can distinguish the similarity of data distribution on different clients.
% It can be seen that the weights of the target client itself, the client with the same label, and the client with the similar label increase as the aining progresses. The weight of the client with the same label is greater than the client with similar classes.   
% One of the 8 clients has the same all kinds of classes as the target client, and another client has the same 3 kinds of classes with the target client.% We record the weight value of each layer on the target client during the training process. 
% there are 8 clients with random 4 kinds of classes on each client. 
% The weight change of one layer on the target client is recorded during the whole training process.
% One of the 8 clients has the same all kinds of classes as the target client, and another client has the same 3 kinds of classes with the target client. 
% The result in Fig.~\ref{fig:weight} shows the changes of the aggregated weights on the target client during the prior, middle and last period of training. It can be seen that the weights of the target client itself, the client with the same label, and the client with the similar label increase as the training progresses. The weight of the client with the same label is greater than the client with similar classes.
%
We also conduct experiments to visualize the relationship between the aggregation weights and the data similarities among clients. We consider 8 clients assigned with ID from 0 to 7, all have four classes data. The data similarities among all clients are emulated by assigning clients of adjacent IDs with similar classes, e.g., client 1 has 4 classes data, while client 2 has three same and one different classes with client 1, and client 3 has three same and one different classes with client 2, and so on. Figure~\ref{fig:weight_visualization} shows the heatmap of the inter-weights among all 8 clients of a certain layer. It can be seen that the weights among close clients with consecutive IDs, i.e., with more overlapping classes, are larger than those of the distant clients, and the highlighted diagonal line shows that the self-weights of each client have the highest values, which further verify that pFedLA can exploit the inter-similarities among heterogeneous clients.

\subsection{Analysis of Communication Efficiency}\label{subsec:communication}
\begin{figure}[t]
\centering
\subcaptionbox{EMNIST\label{1}}{\includegraphics[width = 0.495\linewidth]{ 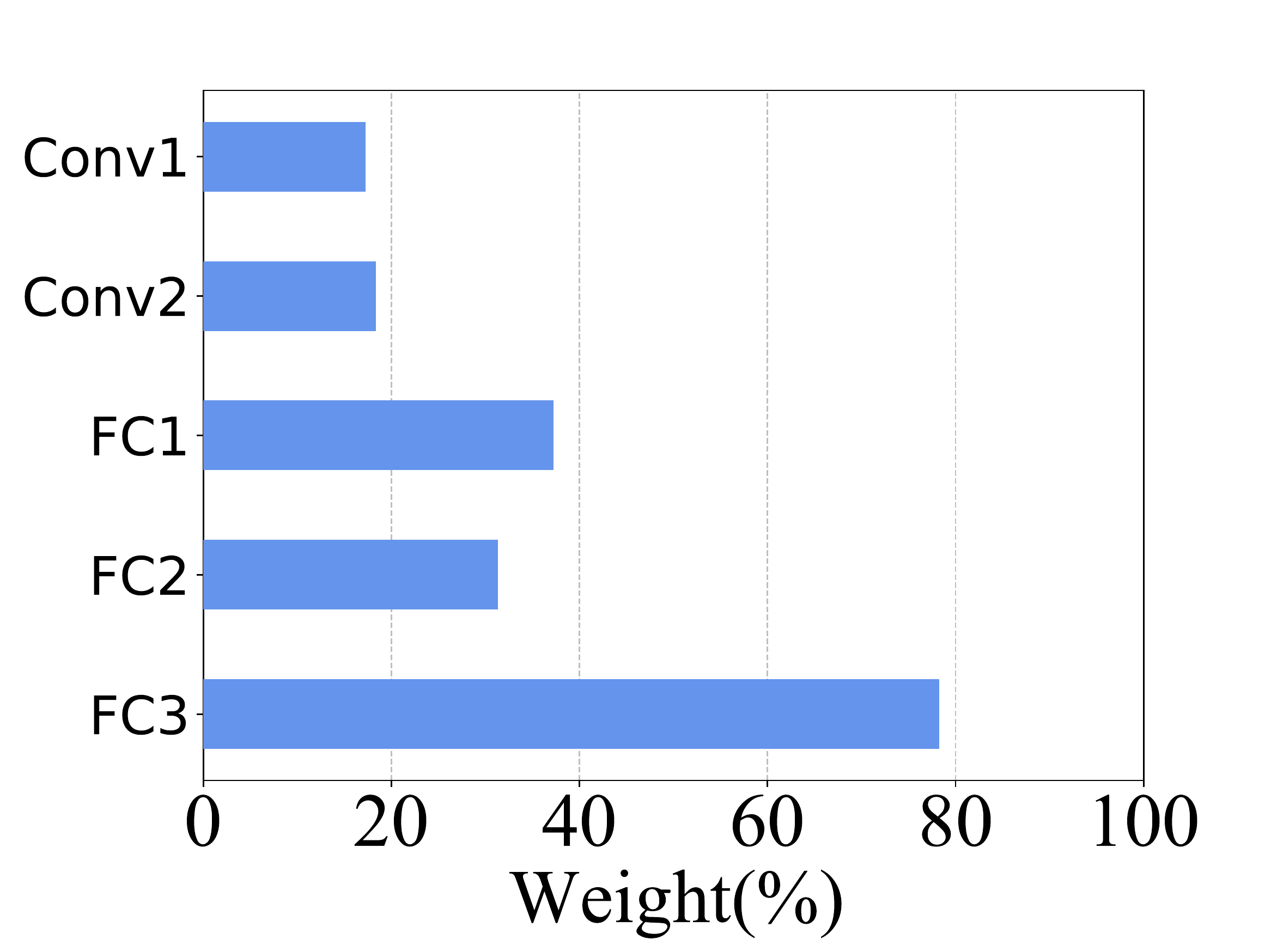}}
\subcaptionbox{CIFAR10\label{2}}{\includegraphics[width = 0.495\linewidth]{ 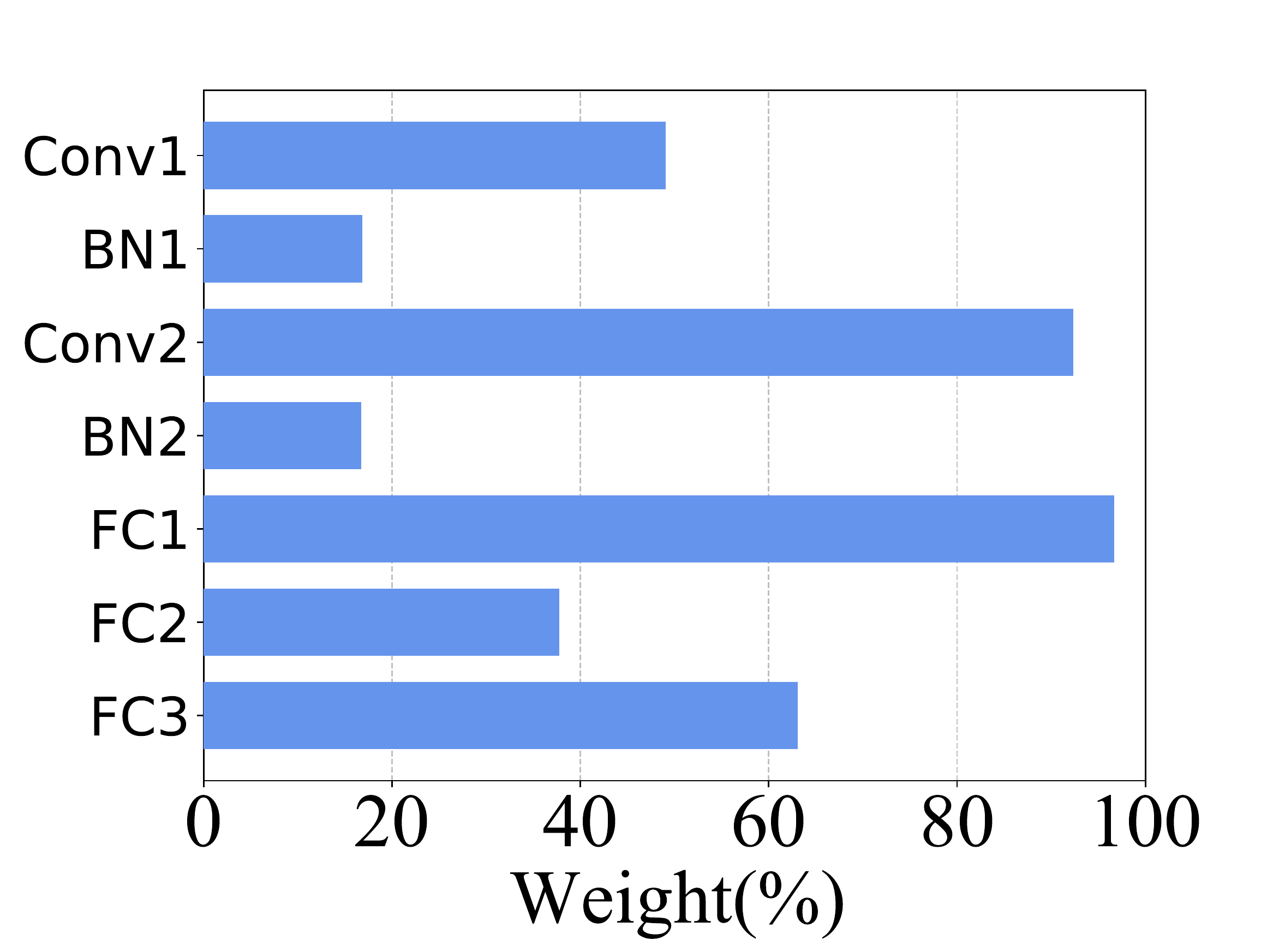}}
\caption{The aggregation weights of all layers for the target client.}
\label{fig:7}
\vspace{-0.3cm}
\end{figure}

% \vspace{-3pt}
In this section, we show the performance of the proposed HeurpFedLA. Table~\ref{tab:heurpfedla} shows the average model accuracy and communication overhead when retaining different local layers that would be absent from the aggregation process. We consider 10 clients with $100\%$ participation over the datasets EMNIST and CIFAR10. 
%%Jie
The aggregation weights of all layers for a target client are shown in Figure~\ref{fig:7}.
For the CIFAR10 dataset, the weights of the first fully-connected layer have the highest values, so the model accuracy performance will be compromised if retaining some layers at local, although the communication overhead can be reduced greatly. For EMNIST dataset, what's different is that the classifier layer has the largest weights, it is observed that the average model accuracy can even increase when retaining some local layers. Such conclusion can also be found in a state-of-the-art work, FedRep \cite{DBLP:conf/icml/CollinsHMS21}, which indicates that removing the classifier layer from the aggregation process can improve the model performance over non-IID datasets.   
It can be explained intuitively that reserving some local layers can avoid the irrelevant knowledge transfer from other clients during the aggregation process.

% \begin{table}[t]
% \centering
% \setlength{\tabcolsep}{1.4mm}{
% \caption{The effect of K.}
% \begin{tabular}{lcccc}
% \toprule
%  {\qquad $K$} & 4  & 5 & 7  & 9
% \\ \cmidrule(lr){1-5}
% CIFAR10 (\%) & 72.23  & 73.15 & \textbf{74.27}  & 71.46  \\ 
% EMNIST (\%) & 95.25  & \textbf{96.34}  & 93.99  & 92.90  \\
% \bottomrule
% \label{tab:tabel_effect_k}
% % \vspace{-0.4cm}
% \end{tabular}}
% \end{table}

\subsection{Effect of $k$}
%% JIe
Different $k$ values are applied to show the effect of retaining local layers. 
% We investigate the effect of $k$ value on learning behavior.
%
Table~\ref{tab:heurpfedla} shows that if retaining different number of the top $k$ layers, the model accuracy will not be affected significantly, which means that HeurpFedLA can apply different $k$ values according to the available communication bandwidth for transmitting the parameters during the pFL iteration, i.e., to do a trade-off between the training efficiency and the communication costs.  

% we can following the way of HeurpFedLA to reduce the communication costs of transmitting the parameters during the pFL iteration. 
% %% JIe
% In real life scenarios, the value of $k$ can be reasonably established according to the requirement of model accuracy and available communication resources, i.e., bandwidth. 

% In the case of non-IID distribution of data between clients, aggregating too many clients’  parameters may cause some problems. i.e., if the data classes on two clients are completely different, aggregating the parameters of these two may even cause a decrease in accuracy. In addition, aggregating too many clients parameters may also cause the model's convergence speed to decrease or even fail to converge. In order to avoid the above problems, in the scenario with more clients, for each client, we only aggregate the K clients with the largest weight. The larger the $K$ value, the more the number of clients involved in aggregation, the more knowledge the client learns, but the greater the possibility of being negatively affected. The Table~\ref{tab:tabel_effect_k} shows the final accuracy results for different K values. The number of participating clients is 100 and the participation rate is $10\%$. It can be seen that the best K-values appear in the moderate position.

\section{Conclusion}\label{sec:conclusion}
In this paper, we have proposed a novel pFL training framework called pFedLA, to achieve personalized model aggregation in a layer-wised aggregation manner. 
% Specifically, instead of calibrating the inter-layer distance for different clients, we have parameterized the layer weights and applied a hypernetwork to learn the user similarity during the pFL training procedure. 
It is shown that such layer-wised aggregation can progressively reinforce the collaboration among similar clients and generate adequate personalization over non-IID datasets that outperform conventional model-wised approaches. In addition, we have provided an improved version of pFedLA that can reduce the communication overhead during the training process with negligible performance loss, and thus can be adapted to large scale FL scenarios where the communication capacity is often limited. 
Extensive evaluations on four different classification tasks demonstrate the feasibility and superior performance of the proposed pFedLA framework. 
% We have also conducted extensive experiments over four different classification tasks, i.e., EMNIST, FashionMNIST, CIFAR10 and CIFAR100, and two different non-IID data distribution settings, and have demonstrated the feasibility and superior performance of the proposed pFedLA framework. 

\section*{Acknowledgements}
% This research was supported by the funding from Hong Kong RGC Research Impact Fund (RIF) with the Project No. R5060-19, General Research Fund (GRF) with the Project No. 152221/19E and 15220320/20E, the National Natural Science Foundation of China (61872310), and Shenzhen Science and Technology Innovation Commission (R2020A045).

% This research was supported by the funding from Hong Kong RGC Research Impact Fund (RIF) with the Project No. R5060-19, General Research Fund (GRF) with the Project No. 152221/19E, 152203/20E, and 152244/21E, the National Natural Science Foundation of China (61872310), and Shenzhen Science and Technology Innovation Commission (JCYJ20200109142008673).
This research was supported by fundings from the Key-Area Research and Development Program of Guangdong Province (No. 2021B0101400003),  Hong Kong RGC Research Impact Fund (No. R5060-19), General Research Fund (No. 152221/19E, 152203/20E, and 152244/21E), the National Natural Science Foundation of China (61872310), and Shenzhen Science and Technology Innovation Commission (JCYJ20200109142008673).

% euristic Improvement

% pFedLA can progressively reinforce the collaboration among similar clients via the layer-wised aggregation  
% %In addition, we have also considered an alternative methods 
% %
% % of each layer to the target client based on similarity metric, we focus on parameterizing each layer's weight via multiple hypernetworks in the server, and alternatively training them with model parameters. 
% %
% By breaking the convention of aggregating the whole model parameters at equivalent weights, this novel personalized aggregation scheme guarantees a flexible and accurate training strategy for personalized Federated Learning. Furthermore, to improve the communication efficiency in large-scale systems, we provide an improved version of our proposed methods dubbed HeurpFedLA to aggregate only the top $k$ layers with highest weights.
% Our empirical study on four different classification tasks (i.e., EMNIST, FashionMNIST, CIFAR10 and CIFAR100) and two different non-IID data distribution cases demonstrates the efficiency of pFedLA and HeurpFedLA. 

%%%%%%%%% REFERENCES
{\small
\bibliographystyle{ieee_fullname}
\bibliography{egbib}
}

\renewcommand{\arraystretch}{1.2}

\end{document}